\pdfoutput=1

\documentclass[11pt,twoside,a4paper,cmspaper,final,collab]{cms-tdr}
\def\svnVersion{a6b4b85-D}\def\svnDate{2020/10/21}\def\cmsCernNoTag{CERN-EP-2020-107}\def\cmsCernDate{\today}\def\cmsMessage{"Published in the Journal of High Energy Physics as \href{http://dx.doi.org/10.1007/JHEP12(2020)085}{\doi{10.1007/JHEP12(2020)085}.}"}

\begin{document}\cmsNoteHeader{HIG-19-003}

\hyphenation{had-ron-i-za-tion}
\hyphenation{cal-or-i-me-ter}
\hyphenation{de-vices}
\newlength\cmsSmallFigWidth
\newlength\cmsBigFigWidth
\newlength\cmsTabWidth
\newlength\cmsTabSkip
\setlength{\cmsTabSkip}{1ex}
\setlength\cmsSmallFigWidth{0.43\textwidth}
\setlength\cmsBigFigWidth{0.7\textwidth}
\setlength\cmsTabWidth{\textwidth}
\providecommand{\cmsTable}[1]{\resizebox{\textwidth}{!}{#1}}
\providecommand{\NA}{\ensuremath{\text{---}}}
\newcommand{\Vjets}{\ensuremath{\PV}+jets\xspace}
\newcommand{\mH}{\ensuremath{m_{\PH}}\xspace}
\newcommand{\mSD}{\ensuremath{m_{\mathrm{SD}}}\xspace}
\newcommand{\nddt}{\ensuremath{N_{2}^{1\mathrm{,DDT}}}\xspace}
\newcommand{\muVal}{\ensuremath{1.9}\xspace}
\newcommand{\muStatErr}{\ensuremath{0.5}\xspace}
\newcommand{\muErrLo}{\ensuremath{0.7}\xspace}
\newcommand{\muErrHi}{\ensuremath{0.9}\xspace}
\newcommand{\muSystErrLo}{\ensuremath{0.3}\xspace}
\newcommand{\muSystErrHi}{\ensuremath{0.3}\xspace}
\newcommand{\muThyErrLo}{\ensuremath{0.3}\xspace}
\newcommand{\muThyErrHi}{\ensuremath{0.6}\xspace}
\newcommand{\muObsLimit}{\ensuremath{3.4}\xspace}
\newcommand{\muExpLimit}{\ensuremath{1.4}\xspace}
\newcommand{\muObsSig}{\ensuremath{2.9}\xspace}
\newcommand{\muExpSig}{\ensuremath{1.7}\xspace}
\newcommand{\muObsLimitMINLO}{\ensuremath{6.4}\xspace}
\newcommand{\muExpLimitMINLO}{\ensuremath{2.9}\xspace}
\newcommand{\muValMINLO}{\ensuremath{3.7}\xspace}
\newcommand{\muStatErrMINLO}{\ensuremath{1.2}\xspace}
\newcommand{\muErrLoMINLO}{\ensuremath{1.5}\xspace}
\newcommand{\muErrHiMINLO}{\ensuremath{1.6}\xspace}
\newcommand{\muSystErrLoMINLO}{\ensuremath{0.7}\xspace}
\newcommand{\muSystErrHiMINLO}{\ensuremath{0.8}\xspace}
\newcommand{\muThyErrLoMINLO}{\ensuremath{0.5}\xspace}
\newcommand{\muThyErrHiMINLO}{\ensuremath{0.8}\xspace}
\newcommand{\muObsSigMINLO}{\ensuremath{2.5}\xspace}
\newcommand{\muObsSigMINLOMuOne}{\ensuremath{1.9}\xspace}
\newcommand{\muExpSigMINLO}{\ensuremath{0.7}\xspace}
\newcommand{\xsecThyVal}{\ensuremath{17.4}\xspace}
\newcommand{\xsecThyErrHi}{\ensuremath{1.5}\xspace}
\newcommand{\xsecThyErrLo}{\ensuremath{2.0}\xspace}
\newcommand{\xsecVal}{\ensuremath{61}\xspace}
\newcommand{\xsecErrLo}{\ensuremath{49}\xspace}
\newcommand{\xsecErrHi}{\ensuremath{50}\xspace}
\newcommand{\xsecStatErr}{\ensuremath{XX}\xspace}
\newcommand{\xsecSystErrLo}{\ensuremath{XX}\xspace}
\newcommand{\xsecSystErrHi}{\ensuremath{XX}\xspace}
\newcommand{\muZVal}{\ensuremath{1.01}\xspace}
\newcommand{\analysisLumi}{\ensuremath{137\fbinv}\xspace}
\newcommand{\muZStatErr}{\ensuremath{0.05}\xspace}
\newcommand{\muZThyErrLo}{\ensuremath{0.09}\xspace}
\newcommand{\muZThyErrHi}{\ensuremath{0.13}\xspace}
\newcommand{\muZSystErrLo}{\ensuremath{0.15}\xspace}
\newcommand{\muZSystErrHi}{\ensuremath{0.20}\xspace}
\newcommand{\muZErrLo}{\ensuremath{0.20}\xspace}
\newcommand{\muZErrHi}{\ensuremath{0.24}\xspace}
\newcommand{\Rpf}{\ensuremath{R_{\mathrm{p}/\mathrm{f}}}\xspace}
\newcommand{\HJMINLO} {{\textsc{HJ-MiNLO}}\xspace}
\newcommand{\MINLO} {{\textsc{MiNLO}}\xspace}
\newcommand{\fxfx} {{\textsc{FxFx}}\xspace}
\newcommand{\ggH}{\ensuremath{\Pg\Pg\PH}\xspace}

\cmsNoteHeader{HIG-19-003}

\title{Inclusive search for highly boosted Higgs bosons decaying to bottom quark-antiquark pairs in proton-proton collisions at \texorpdfstring{$\sqrt{s} = 13\TeV$}{sqrt(s)=13 TeV}}

\date{\today}

\abstract{
A search for standard model Higgs bosons (\PH) produced with transverse momentum (\pt) greater than 450\GeV and decaying to bottom quark-antiquark pairs ($\bbbar$) is performed using proton-proton collision data collected by the CMS experiment at the LHC at $\sqrt{s}=13\TeV$.
The data sample corresponds to an integrated luminosity of \analysisLumi.
The search is inclusive in the Higgs boson production mode.
Highly Lorentz-boosted Higgs bosons decaying to $\bbbar$ are reconstructed as single large-radius jets, and are identified using jet substructure and a dedicated \PQb tagging technique based on a deep neural network.
The method is validated with $\PZ\to\bbbar$ decays.
For a Higgs boson mass of 125\GeV, an excess of events above the background assuming no Higgs boson production is observed with a local significance of \muObsSigMINLO standard deviations ($\sigma$), while the expectation is \muExpSigMINLO.
The corresponding signal strength and local significance with respect to the standard model expectation are $\mu_{\PH} = \muValMINLO\pm\muStatErrMINLO\stat_{-\muSystErrLoMINLO}^{+\muSystErrHiMINLO}\syst_{-\muThyErrLoMINLO}^{+\muThyErrHiMINLO}\thy$ and $\muObsSigMINLOMuOne\,\sigma$.
Additionally, an unfolded differential cross section as a function of Higgs boson \pt for the gluon fusion production mode is presented, assuming the other production modes occur at the expected rates.
}

\hypersetup{
pdfauthor={CMS Collaboration},
pdftitle={Inclusive search for highly boosted Higgs bosons decaying to bottom quark-antiquark pairs in proton-proton collisions at sqrt(s)= 13 TeV},
pdfsubject={CMS},
pdfkeywords={CMS, physics, boosted Higgs, bottom quark-antiquark decay, gluon fusion}}

\maketitle

\section{Introduction}
\label{sec:intro}

The observation of a new boson consistent with the standard model (SM) Higgs boson (\PH) and the subsequent measurements of its properties~\cite{:2012gk,:2012gu,Chatrchyan:2013lba} have advanced the understanding of electroweak (EW) symmetry breaking and the origin of the mass of elementary particles~\cite{Salam:1968rm,Glashow:1961tr,Weinberg:1967tq,PhysRevLett.13.321,Higgs:1964ia,PhysRevLett.13.508,PhysRev.145.1156,PhysRevLett.13.585}.
The \PH boson has been observed at the CERN LHC in all of its main expected production modes and several decay modes, including decays to bottom quark-antiquark pairs ($\bbbar$) when produced in association with a \PW or \PZ boson~\cite{Sirunyan:2018kst,Aaboud:2018zhk}.
Recently, there has been considerable interest in the measurement of Higgs bosons produced with high transverse momentum, \pt, where measurements in the $\PH(\bbbar)$ decay channel have better sensitivity than traditional channels because of its large branching fraction, $\mathcal{B}(\PH\to\bbbar)= 58.1\%$~\cite{deFlorian:2016spz}.
Advances in the identification of large-radius jets~\cite{Seymour:1991cb,Seymour:1993mx,Seymour:1994by,Butterworth:2002tt,Butterworth:2008iy} resulting from massive color singlet particles with large transverse momentum and decaying to $\bbbar$ pairs have improved the sensitivity of this channel, as demonstrated by the CMS~\cite{Sirunyan:2017ezt,CMS-DP-2018-046} and ATLAS~\cite{Aad:2019uoz} Collaborations.
The first search for high-\pt $\PH(\bbbar)$ events by the CMS Collaboration~\cite{Sirunyan:2017dgc} demonstrated the experimental sensitivity of this channel, with an expected significance of 0.7 standard deviations ($\sigma$) based on a different theoretical expectation than the latest one used in this paper.
Measurements of high-\pt $\PH(\bbbar)$ events provide an alternative approach to study the top quark Yukawa coupling, complementary to associated \PH production with a top quark-antiquark pair ($\ttbar\PH$), and may be sensitive to effects from physics beyond the SM~\cite{Sirunyan:2018sgc,Grojean:2013nya,Dawson:2015gka,Schlaffer:2014osa,Grazzini:2017szg,Grazzini:2016paz,Bishara:2016jga,Li:2019pag}.
At the highest \pt, this measurement can resolve loop-induced contributions to the \ggH process from new particles, such as a top quark partner, which would be described by an effective \ggH vertex at low \pt.

This paper reports the results of an inclusive search for high-\pt Higgs bosons decaying to $\bbbar$ pairs in proton-proton ($\Pp\Pp$) collisions at $\sqrt{s} = 13\TeV$.
The data set, collected with the CMS detector at the LHC in 2016--2018, corresponds to an integrated luminosity of \analysisLumi.
The search is inclusive in the Higgs boson production mode.
The highly Lorentz-boosted $\PH(\bbbar)$ candidates are reconstructed as single large-radius jets with the jet mass consistent with that of the observed Higgs boson~\cite{Butterworth:2008iy}.
The candidate jet is required to have $\pt>450\GeV$ to satisfy restrictive trigger requirements that suppress the large background from jets produced via the strong interaction, referred to as quantum chromodynamics (QCD) multijet events.
To further distinguish the \PH candidates from the background, the jet is required to have a two-prong substructure, as well as displaced tracks and decay vertices consistent with the $\PH(\bbbar)$ signal, identified with a dedicated algorithm that detects the presence of \PQb hadrons in the jet (\PQb tagging).
The events are divided into six adjacent \pt categories.
The background from QCD multijet production is difficult to model parametrically, and it is therefore estimated in data by relating the event yields in the signal region to those in a control region defined by inverting the \PQb tagging requirement, which is designed to have reduced correlation with jet mass and \pt.
The presence of the \PW and \PZ boson resonances in the jet mass distribution is used to constrain various systematic uncertainties and to validate the analysis.
A separate control region is used to improve the modeling of the \ttbar background.
A simultaneous fit to the distributions of the jet mass in all \pt categories is performed to determine the normalizations and shapes of the jet mass distributions for the backgrounds and to extract the inclusive $\PH(\bbbar)$ signal strength with respect to the SM expectation.
The differential cross section for the \ggH Higgs boson \pt is also extracted under the assumption that \PH production through other modes occurs at the SM rate.

In contrast with the previous CMS result, the Higgs boson \pt spectrum from \ggH production is modeled with the \HJMINLO generator~\cite{Hamilton:2012rf,Becker:2669113,Neumann:2018bsx}, which includes effects of the finite top quark mass effects to higher order in QCD.
The predicted cross section is compatible with the latest theoretical calculations~\cite{Jones:2018hbb,Lindert:2018iug}, and is smaller than that used previously~\cite{Sirunyan:2017dgc}.
Another major improvement is the development of a \PQb tagging algorithm based on a deep neural network with better $\PH(\bbbar)$ signal efficiency.

This paper is organized as follows.
A brief description of the CMS detector is given in Section~\ref{sec:detector}.
Section~\ref{sec:samples} provides a summary of the various simulated samples used in the analysis.
Section~\ref{sec:physobj} describes the event reconstruction and selection criteria used to define the signal and control regions.
The background estimation methods are detailed in Section~\ref{sec:bkgest}.
Section~\ref{sec:unc} lists the sources of systematic uncertainty and their statistical treatment.
Section~\ref{sec:results} describes the statistical procedure used to derive the results, and reports the results in terms of signal strength modifiers and differential cross sections.
Finally, the results are summarized in Section~\ref{sec:summary}.

\section{The CMS detector}
\label{sec:detector}
The central feature of the CMS apparatus is a superconducting solenoid of 6\unit{m} internal diameter, providing a magnetic field of 3.8\unit{T} inside its volume.
Within the solenoid volume are a silicon pixel and strip tracker, a lead tungstate crystal electromagnetic calorimeter, and a brass and scintillator hadron calorimeter, each composed of a barrel and two endcap sections.
Forward calorimeters extend the pseudorapidity ($\eta$) coverage provided by the barrel and endcap detectors.
Muons are detected in gas-ionization chambers embedded in the steel flux-return yoke outside the solenoid.

Events of interest are selected using a two-tiered trigger system~\cite{Khachatryan:2016bia}.
The first level, composed of custom hardware processors, uses information from the calorimeters and muon detectors to select events at a rate of around 100\unit{kHz} within a time interval of less than 4\mus.
The second level, known as the high-level trigger, consists of a farm of processors running a version of the full event reconstruction software optimized for fast processing, and reduces the event rate to around 1\unit{kHz} before data storage.

A more detailed description of the CMS detector, together with a definition of the coordinate system used and the relevant kinematic variables, can be found in Ref.~\cite{Chatrchyan:2008zzk}.

\section{Simulated samples}
\label{sec:samples}

Simulated samples of signal and background events are produced using various Monte Carlo (MC) event generators, with the CMS detector response modeled by \GEANTfour~\cite{GEANT4}.

For 2016 running conditions, the QCD multijet and $\PZ$+jets processes are modeled at leading order (LO) accuracy using the \MGvATNLO~v2.2.2 generator~\cite{Alwall:2014hca}.
The $\PW$+jets process is modeled at LO accuracy with \MGvATNLO~v2.3.3.
The vector boson (\PV) samples include decays of the bosons to all flavors of quarks, $\PV(\qqbar)$, and include up to 3 (4) extra partons at the matrix element level for $\PW$+jets ($\PZ$+jets).
Jets from the matrix element calculation and the parton shower description are matched using the MLM prescription~\cite{Alwall:2007fs}.
The \ttbar and single top quark processes are modeled at next-to-LO (NLO) using \POWHEG~2.0~\cite{Nason:2004rx,Frixione:2007vw,Alioli:2010xd,Frixione:2007nw,Frederix:2012dh,Re:2010bp}.
Diboson processes are modeled at LO accuracy with \PYTHIA~8.205~\cite{Sjostrand:2014zea}.

{\tolerance=1000
For 2017 and 2018 running conditions, the same configurations are used, but with newer generator versions.
The QCD multijet and \Vjets processes are modeled using \MGvATNLO~v2.4.2, and the diboson processes are modeled with \PYTHIA~8.226.
\par}

For all years, the cross sections for the \Vjets samples are corrected as functions of boson \pt for higher-order QCD and EW effects.
The QCD NLO corrections are derived using \MGvATNLO, simulating \PW and \PZ production with up to 2 additional partons and \fxfx matching to the parton shower~\cite{Frederix:2012ps}.
The EW NLO corrections are taken from theoretical calculations in Ref.~\cite{Kallweit:2014xda,Kallweit:2015dum,Kallweit:2015fta,Lindert:2017olm}.
Additionally, the total cross sections for the diboson samples are corrected to next-to-NLO (NNLO) accuracy with the \MCFM~7.0 program~\cite{Campbell:2010ff}.

The \ggH production process is simulated using the {\HJMINLO}~\cite{Frixione:2007vw,Luisoni:2013kna,Hamilton:2012rf,Becker:2669113} event generator with mass $m_{\PH} = 125\GeV$ and including finite top quark mass effects, following the recommendation in Ref.~\cite{Becker:2669113}.
Additionally, a sample of \ggH events is generated with \POWHEG~\cite{Bagnaschi2012} and corrected for the effects of the finite top quark mass using the same procedure as described in Ref.~\cite{Sirunyan:2017dgc}, where the NLO to LO ratio of the \pt spectrum is approximated by expanding in powers of the inverse square of the top quark mass.
The \POWHEG generator is used to model Higgs boson production through vector boson fusion (VBF), $\PV\PH$ associated production, and $\ttbar\PH$ channels~\cite{Nason:2009ai,Luisoni:2013kna,Hartanto:2015uka}.
The \pt spectrum of the Higgs boson for the VBF production mode is re-weighted to account for next-to-NNLO corrections to the cross section~\cite{Cacciari:2015jma,Dreyer:2016oyx}.
These corrections have a negligible effect on the yield for this process for events with Higgs boson $\pt>450\GeV$.

For parton showering and hadronization, the \POWHEG and \MGvATNLO samples are interfaced with \PYTHIA~8.205 (8.230) for 2016 (2017 and 2018) running conditions.
The \PYTHIA parameters for the underlying event description are set to the CUETP8M1~\cite{Khachatryan:2015pea} (CP5~\cite{Sirunyan:2019dfx}) tune, except for the \ttbar sample for 2016, which uses the CUETP8M2T4 tune~\cite{CMS-PAS-TOP-16-021}.
For 2016 samples, the parton distribution function set {NNPDF3.0}~\cite{Ball:2014uwa} is used, with the accuracy (LO or NLO) corresponding to that used in the matrix element calculations, while for 2017 and 2018 samples, {NNPDF3.1}~\cite{Ball:2017nwa} at NNLO accuracy is used for all processes.

\section{Event reconstruction and selection}
\label{sec:physobj}

Event reconstruction is based on a particle-flow algorithm~\cite{Sirunyan:2017ulk}, which aims to reconstruct and identify each individual particle with an optimized combination of information from the various elements of the CMS detector.
The algorithm identifies each reconstructed particle as an electron, a muon, a photon, or a charged or neutral hadron.
The missing transverse momentum vector is defined as the negative vector sum of the transverse momenta of all the particles identified in the event, and its magnitude is referred to as $\ptmiss$.
The candidate vertex with the largest value of summed physics-object $\pt^2$ is taken to be the primary $\Pp\Pp$ interaction vertex.
The physics objects are the jets, clustered using the jet finding algorithm~\cite{Cacciari:2008gp} with the tracks assigned to candidate vertices as inputs, and the associated missing transverse momentum, taken as the negative vector sum of the \pt of those jets.

Particles are clustered into jets using the anti-\kt algorithm with a distance parameter of 0.8 (AK8 jets) or 0.4 (AK4 jets).
The larger radius of the AK8 jet better captures the decay products of the high-\pt $\PH(\bbbar)$ signal.
The clustering algorithms are implemented by the \FASTJET package~\cite{Cacciari:2011ma}.
To mitigate the effect from the contributions of simultaneous $\Pp\Pp$ collisions (pileup), the pileup per-particle identification algorithm~\cite{Bertolini:2014bba,Sirunyan:2020foa} assigns a weight to each particle prior to jet clustering based on the likelihood of the particle to originate from the hard scattering vertex.
Further corrections are applied to the jet energy as a function of jet $\eta$ and \pt to bring the average measured response of jets to that of jets made directly from the generated particles before simulation of the detector response~\cite{Khachatryan:2016kdb}.
These corrections are derived separately for each data collection year.
Jet identification criteria are applied to remove spurious jets associated with calorimeter noise as well as those associated with muon and electron candidates that are either misreconstructed
or isolated.
Specifically, jets are required to have neutral hadron and photon energy fractions less than 90\%, nonzero charged hadron energy fractions, muon energy fractions less than 80\%, and at least two constituent particles~\cite{CMS-PAS-JME-16-003}.
Additionally, AK8 jets are rejected if a photon with $\pt > 175\GeV$ is reconstructed within the jet.

A combination of several event selection criteria is used for the event trigger, all of which impose minimum thresholds on either the AK8 jet \pt or the event \HT, defined as the scalar \pt sum of all jets in the event with $\abs{\eta} < 3.0$.
For AK8 jets used in the trigger selection, a minimum threshold is also imposed on the trimmed jet mass~\cite{Krohn:2009th}, where remnants of soft radiation are removed before computing the mass, which allows the $\HT$ or \pt thresholds to be reduced while maintaining manageable trigger rates.
The trigger selection efficiency is greater than 95\% for events with at least one AK8 jet with $\abs{\eta} < 2.5$, mass greater than 47\GeV and $\pt > 450\,(525, 500) \GeV$ for 2016 (2017, 2018) data.

To reduce backgrounds from SM EW processes, events are vetoed if they contain isolated electrons, isolated muons, or hadronically decaying $\tau$ leptons with $\pt > 10$, 10, or 18\GeV and $\abs{\eta} < 2.5$, 2.4, or 2.3, respectively.
For electrons and muons, an isolation variable is calculated as the pileup-corrected \pt sum of the charged hadrons and neutral particles surrounding the lepton divided by the lepton \pt.
For charged particles, only those associated with the primary vertex are considered in the isolation variable.
For neutral particles, the pileup correction consists of subtracting the energy deposited in the isolation cone by charged hadrons not associated with the primary vertex, multiplied by a factor of $0.5$.
This factor corresponds approximately to the ratio of neutral to charged hadron production in pileup interactions~\cite{Chatrchyan:2012vp}.
The isolation variable for electrons and muons is required to be less than 15 or 25\%, respectively, depending on $\eta$~\cite{Khachatryan:2015hwa,Sirunyan:2018fpa}.

For each event, the leading AK8 jet in \pt is selected to be the $\PH(\bbbar)$ candidate, which is around 60\% efficient for the $\ggH$ production mode.
Alternative $\PH(\bbbar)$ candidate jet selection criteria were considered, but were not found to improve the sensitivity.
The AK8 jet is required to have $\abs{\eta}<2.5$.
To reduce the top quark contamination, events are vetoed if they have $\ptmiss >140\GeV$, or if they contain a \cPqb-tagged~\cite{Sirunyan:2017ezt} AK4 jet with $\pt > 30\GeV$ located in the opposite hemisphere from the leading AK8 jet ($\Delta\phi(\text{AK4},\text{AK8}) > \pi/2$).
The chosen threshold for the AK4 jet \cPqb-tagging algorithm corresponds to a 1\% probability to misidentify a jet arising from a light flavor quark or gluon and a 77\% probability to correctly identify a jet arising from a \PQb quark in 2017 detector conditions.
Approximately 60\% of \ttbar events are rejected by this selection.

The soft-drop (SD) algorithm~\cite{Larkoski:2014wba} with angular exponent $\beta=0$ and soft radiation fraction $z=0.1$ is applied to the Higgs boson jet candidate to remove soft and wide-angle radiation.
The parameter $\beta$ controls the grooming profile as a function of subjet separation; for $\beta=0$, the algorithm is independent of subjet separation, and is equivalent to the modified mass-drop tagger~\cite{Dasgupta:2013ihk}.
The resulting SD jet mass, \mSD, is strongly reduced for background QCD multijet events, where large jet masses arise from wide-angle gluon radiation.
Conversely, the algorithm preserves the mass of jets from heavy boson decays.
Corrections to the \mSD values from simulation are derived from a comparison of simulated and measured samples in a region enriched with merged $\PW(\qqbar)$ decays from \ttbar events~\cite{CMS-PAS-JME-16-003}.
The \mSD corrections remove a residual dependence on the jet \pt, and match the simulated jet mass scale and resolution to those observed in data.

The resulting \mSD distributions are binned from 47 to 201\GeV with a bin width of 7\GeV.
The lower bound is sufficiently above the trigger threshold to be insensitive to differences between the online and offline mass calculations, and the bin width corresponds to the \mSD resolution near the $\PV$ resonances.
The dimensionless mass scale variable for QCD multijet jets, $\rho(\mSD,\pt)=2\ln(\mSD/\pt)$~\cite{Dasgupta:2013ihk,Dolen:2016kst}, is used to characterize the correlation between the jet \PQb tagging discriminator, jet mass, and jet \pt.
Its distribution is roughly invariant in different ranges of jet \pt.
For each \pt category, only those \mSD bins that satisfy
\begin{equation}
  \label{eqn:rhocut}
  \rho(0.5\mSD^{\mathrm{lo}} + 0.5\mSD^{\mathrm{up}},\, 0.7\pt^{\mathrm{lo}} + 0.3\pt^{\mathrm{up}}) < -2.1
\end{equation}
are considered, where $\mSD^{\mathrm{up}}$ ($\pt^{\mathrm{up}}$) is the upper \mSD (\pt) bound and $\mSD^{\mathrm{lo}}$ ($\pt^{\mathrm{lo}}$) is the lower \mSD (\pt) bound.
In this restriction, the lower \pt bound is weighted more heavily because of the steeply falling QCD multijet \pt distribution.
This upper bound on $\rho$ is imposed to avoid instabilities at the edges of the distribution due to finite cone limitations from the jet clustering.
This requirement is about 98\% efficient for the $\PH(\bbbar)$ signal.

The $N_2^{1}$ variable~\cite{Moult:2016cvt} is used to determine how consistent a jet is with having a two-prong substructure.
It is based on a ratio of 2-point ($_{1}e_{2}$) and 3-point  ($_{2}e_{3}$) generalized energy correlation functions~\cite{Larkoski:2013eya}:
\begin{equation}
  \begin{aligned}
_{1}e_{2} &= \sum_{1\leq i < j \leq n} z_{i}z_{j}\Delta R_{ij} , \\
_{2}e_{3} &= \sum_{1\leq i < j < k\leq n} z_{i}z_{j}z_{k} \min \{\Delta R_{ij}\Delta R_{ik}, \Delta R_{ij}\Delta R_{jk}, \Delta R_{ik}\Delta R_{jk} \} ,
  \end{aligned}
\end{equation}
where $z_i$ represents the energy fraction of the constituent $i$ in the jet, and $\Delta R_{ij}$ is the angular separation between constituents $i$ and $j$.
These generalized energy correlation functions ${_{v}e_n}$ are sensitive to correlations of $v$ pairwise angles among $n$ jet constituents~\cite{Moult:2016cvt}.
For a two-prong structure, signal jets have a stronger 2-point correlation than a 3-point one.
The discriminant variable $N_2^{1}$ is defined as
\begin{equation}
\quad N_{2}^{1}= \frac{_{2}e_{3}}{(_{1}e_{2})^{2}} .
\end{equation}
The calculation of $N_2^{1}$ is based on the jet constituents after application of the SD grooming algorithm to the jet.
It provides excellent discrimination between two-prong signal jets and QCD background jets.
However, imposing requirements on $N_2^{1}$, or other similar variables, distorts the jet mass distributions differently depending on the jet \pt~\cite{Thaler:2010tr}.
To minimize this distortion, a transformation is applied to $N_{2}^{1}$ following the designed decorrelated tagger technique~\cite{Dolen:2016kst}, reducing its correlation with $\rho$ and \pt in multijet events.
The transformed variable is defined as $\nddt \equiv N_2^{1} - X_{(26\%)}$, where $X_{(26\%)}$ is the value corresponding to the 26th percentile of the $N_2^{1}$ distribution in simulated QCD events, as a function of $\rho$ and \pt.
The transformation is derived in bins of $\rho$ and \pt.
This ensures that the selection $\nddt<0$ yields a constant background efficiency for QCD events across the $\rho$ and \pt range considered in this search.
The chosen efficiency of 26\% maximizes the signal sensitivity.

\begin{figure*}[hbtp]
  \centering
  \includegraphics[width=\cmsBigFigWidth]{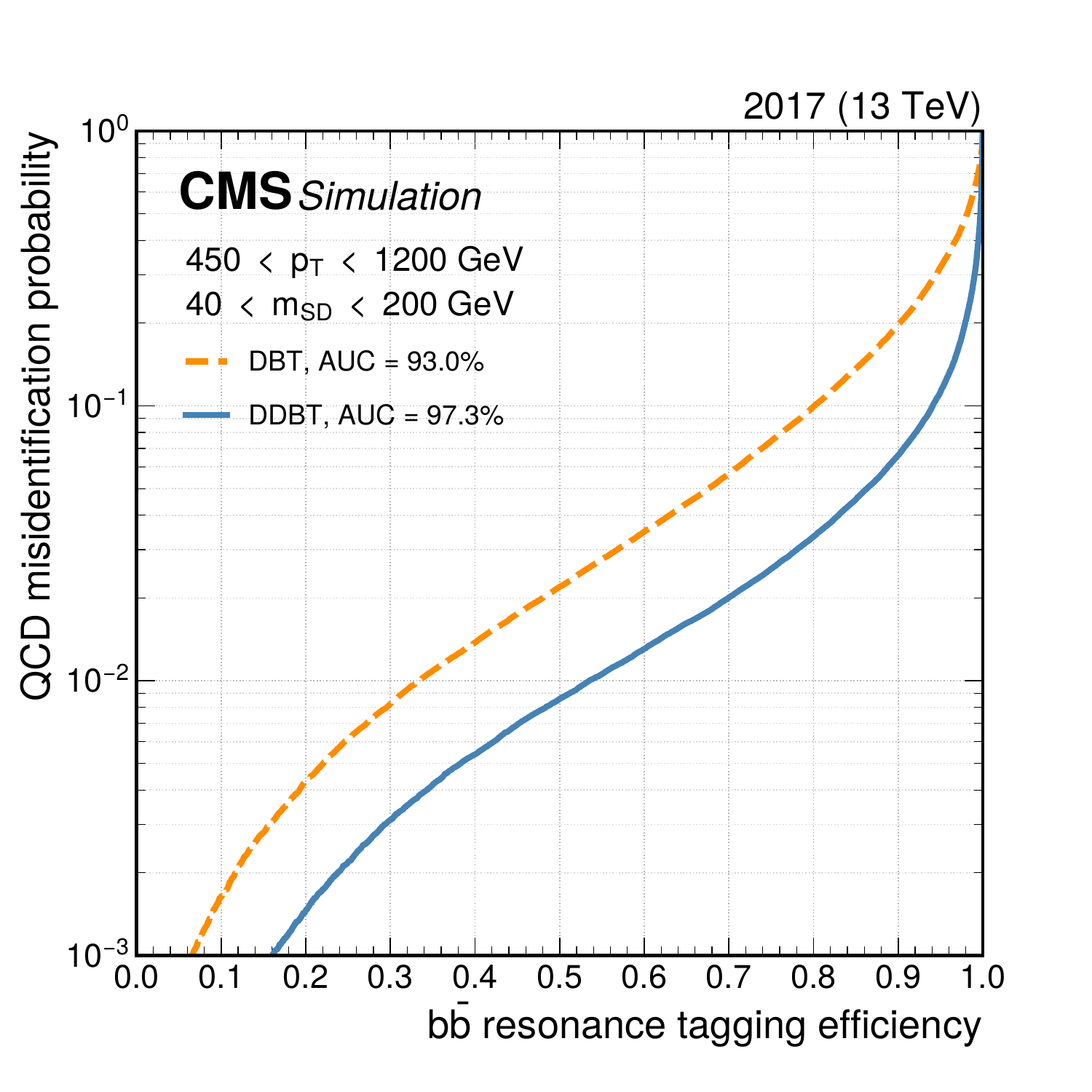}
  \caption{The performance curves of misidentification probability for jets originating from QCD multijet production versus the identification probability for $\bbbar$ resonance jets for the DBT (orange dashed line) used in a prior CMS result and the DDBT (blue solid line). The $\bbbar$ resonances are generated with variable masses in the range 15--250\GeV. The curves are evaluated with simulation corresponding to the detector conditions in 2017. Jets are required to have \pt in the range 450--1200\GeV and \mSD in the range 40--200\GeV. The area under the curve (AUC) is reported as a performance metric for both algorithms.}
      \label{fig:roc}
\end{figure*}

Jets likely to originate from the merging of the fragmentation products of two \PQb quarks are selected using an algorithm based on a deep neural network, composed of multiple layers between input and output, referred to here as the deep double-\PQb tagger (DDBT)~\cite{CMS-DP-2018-046,Sirunyan:2017ezt}.
The algorithm takes as inputs several high-level observables that characterize the distinct properties of \PQb hadrons and their momentum directions in relation to the two subjet candidate axes, as well as low-level track and vertex observables.
Events where the selected AK8 jet is double-\PQb tagged constitute the ``passing,'' or signal, region, while events failing the DDBT form the ``failing'' region, which is used to estimate the QCD multijet background in the signal region.
Specifically, an AK8 jet is considered double-\PQb tagged if its DDBT discriminator value exceeds a threshold corresponding approximately to a 1\% misidentification probability for QCD jets.
This threshold corresponds to a 54\% efficiency for reconstructed scalar boson resonances with variable masses decaying to $\bbbar$ in the range $40 < \mSD < 200\GeV$ and $450 < \pt < 1200\GeV$ in simulation corresponding to the detector conditions in 2017.
The performance of the DDBT algorithm for 2018 detector conditions is approximately the same, while the performance for 2016 ones is slightly worse (45\% efficiency for $\bbbar$ resonances in the same \mSD and \pt range and for the same misidentification probability) because the CMS pixel tracker was upgraded between 2016 and 2017~\cite{Dominguez:1481838}.
Compared to the previous double-\PQb tagger (DBT) algorithm~\cite{Sirunyan:2017ezt} used in a prior CMS result~\cite{Sirunyan:2017dgc}, the DDBT improves the $\bbbar$ tagging efficiency by a factor of about 1.6 for the same detector conditions and QCD misidentification probability.
For SM {\ggH} production specifically, the tagging efficiency is approximately 60\%, an improvement over the previous algorithm by a factor of about 1.3.
Figure~\ref{fig:roc} shows the performance curves of misidentification probability for QCD jets versus the identification probability for $\bbbar$ resonance jets for the previous DBT algorithm and the DDBT algorithm in simulation corresponding to 2017 detector conditions.

After all selections are applied, the Higgs boson candidate jet is categorized into the DDBT passing or failing region, each with 22 \mSD bins evenly dividing the range 47--201\GeV, and split further into six jet \pt categories with bin boundaries of 450, 500, 550, 600, 675, 800, and 1200\GeV.
The upper \pt bound of 1200\GeV does not have a significant impact on the sensitivity and excludes a region where the QCD multijet background is difficult to model.
The remaining \pt binning is optimized for best signal significance, and the upper \mSD bound is due to the requirements imposed on the jet $\rho$.
Specifically, bins that do not satisfy Eq.~(\ref{eqn:rhocut}) are removed, resulting in a total of 124 bins each for the passing and failing regions.
Namely, the upper \mSD bound for the first two \pt categories are 166\GeV and 180\GeV, respectively.
For the Higgs boson signal processes in the DDBT passing region, the dominant production mode is \ggH (56\%), followed by VBF (26\%), $\PV\PH$ (13\%), and $\ttbar\PH$ (5\%).

\section{Background estimation}
\label{sec:bkgest}

The dominant background in the signal region is QCD multijet production.
The \Vjets processes are significant resonant backgrounds.
The \ttbar process constitutes a significant nonresonant background across the \mSD spectrum.
Other EW processes, including diboson, triboson, and $\ttbar\PV$, are estimated from simulation and found to be negligible.

The \Vjets background is modeled using simulation.
Their overall contribution is less than 6\% of the total background in the DDBT passing region.
The normalizations and shapes of the simulated \Vjets background are corrected for NLO QCD and EW effects.

The contribution of \ttbar production to the total background is obtained from simulation, where the normalization and DDBT efficiency are corrected with scale factors derived from a \ttbar-enriched control sample.
The control sample targets semileptonic \ttbar production, consisting of events with an energetic muon with $\pt > 55\GeV$ and $\abs{\eta} < 2.1$, a leading AK8 jet with $\pt > 400\GeV$, and an additional \cPqb-tagged AK4 jet that is separated from the leading AK8 jet by $\Delta R > 0.8$.
The AK8 jet with the highest \pt is taken to be the candidate jet.
Using the same candidate jet requirements that define the signal selection, DDBT passing and failing regions are constructed in both data and simulation.
Due to the relatively low event count in the control sample, the inclusive event counts for $47 < \mSD < 201\GeV$ and $\pt > 400\GeV$ are used, totalling 438 (6301) events in the data passing (failing) region.
The fraction of \ttbar background relative to the total background expected in this control sample is 72\%.
Both the absolute normalization and DDBT efficiency of the \ttbar contribution are allowed to vary without constraint from the simulation expectation, but are forced to vary identically in the \ttbar control region and the signal region in the simultaneous fit, thus constraining the background expectation and DDBT mistag probability for this process.
The net contribution is about 8\% of the total background in the $110 < \mSD < 131\GeV$ range of the DDBT passing region.

The main background in the DDBT passing region, QCD multijet production, has a jet mass shape that depends on \pt and is difficult to model parametrically.
Therefore, we estimate it using the background-enriched failing region, i.e., events failing the DDBT selection, together with a ``pass-fail ratio'' function, \Rpf.
Ideally, \Rpf would be constant as a function of jet mass and \pt, as the DDBT discriminator is designed to be uncorrelated from both variables: the training procedure incorporates a penalty term to the loss function for differences in the jet mass distribution between the passing and failing events, and the training samples are weighted such that the loss function is independent of jet \pt.
Nonetheless, the DDBT exhibits some anticorrelation at high tagger discriminator values and low jet mass, i.e., the mass distributions are different in the passing and failing regions. Additionally, residual differences in \Rpf may arise from discrepancies in tagger performance between data and simulation.
To account for both effects, \Rpf is separated into two components: an expected pass-fail ratio is taken from simulated QCD multijet events by fitting a two-dimensional second-order Bernstein polynomial~\cite{bernstein} in $\rho$ and \pt, $\epsilon^{\mathrm{QCD}}(\rho,\,\pt)$, to the distributions in simulation; and a data residual correction is parametrized using a Bernstein polynomial in $\rho$ and \pt.
The complete pass-fail ratio in data is given by the product of these two factors,
\begin{equation}
\begin{aligned}
\Rpf(\rho,\pt) &= \sum^{n_{\rho}}_{k=0} \sum^{n_{\pt}}_{\ell=0} a_{k, \ell} b_{k, n_{\rho}}(\rho)  b_{\ell, n_{\pt}}(\pt) \epsilon^{\mathrm{QCD}}(\rho, \pt) ,~
\end{aligned}
\end{equation}
where $n_{\rho}$ is the degree of the polynomial in $\rho$, $n_{\pt}$ is the degree of the polynomial in \pt, $a_{k, \ell}$ is a Bernstein coefficient, and
\begin{equation}
\begin{aligned}
b_{{\nu ,n}}(x)&=\binom{n}{\nu} x^{{\nu }}\left(1-x\right)^{{n-\nu}}
\end{aligned}
\end{equation}
is a Bernstein basis polynomial of degree $n$.

The pass-fail ratio \Rpf is determined from a simultaneous binned fit to the \mSD data distributions in the DDBT passing and failing regions across the whole jet mass and \pt range, accounting for the contributions from signal and non-QCD backgrounds.
In this fit, the coefficients $a_{k,\ell}$ (data correction) are fitted with no external constraints, while the $\epsilon^{\mathrm{QCD}}$ coefficients and their associated uncertainties are taken from the separate fit to the QCD simulation.
The \pt bin widths, which vary from 50 to 400\GeV, are chosen to provide enough data points to constrain the shape of \Rpf.
To determine the minimum degree of polynomial necessary to fit the data, a Fisher $F$-test~\cite{ref:ftest} is performed.
As the magnitude of data-to-simulation discrepancies can vary among the data samples and their corresponding simulation samples, an $F$-test is performed independently for each of the three data taking years.
For the 2016 data sample, it is found that a polynomial of order $(n_{\rho},\,n_{\pt})=(2,\,1)$ is needed to provide a sufficient goodness of fit with respect to increased orders ($p>0.05$), while for 2017 and 2018 data, a residual polynomial of order $(n_{\rho},\,n_{\pt})=(1,\,1)$ is found to be sufficient.

\begin{figure*}[hbtp]
  \centering
      \includegraphics[width=\cmsBigFigWidth]{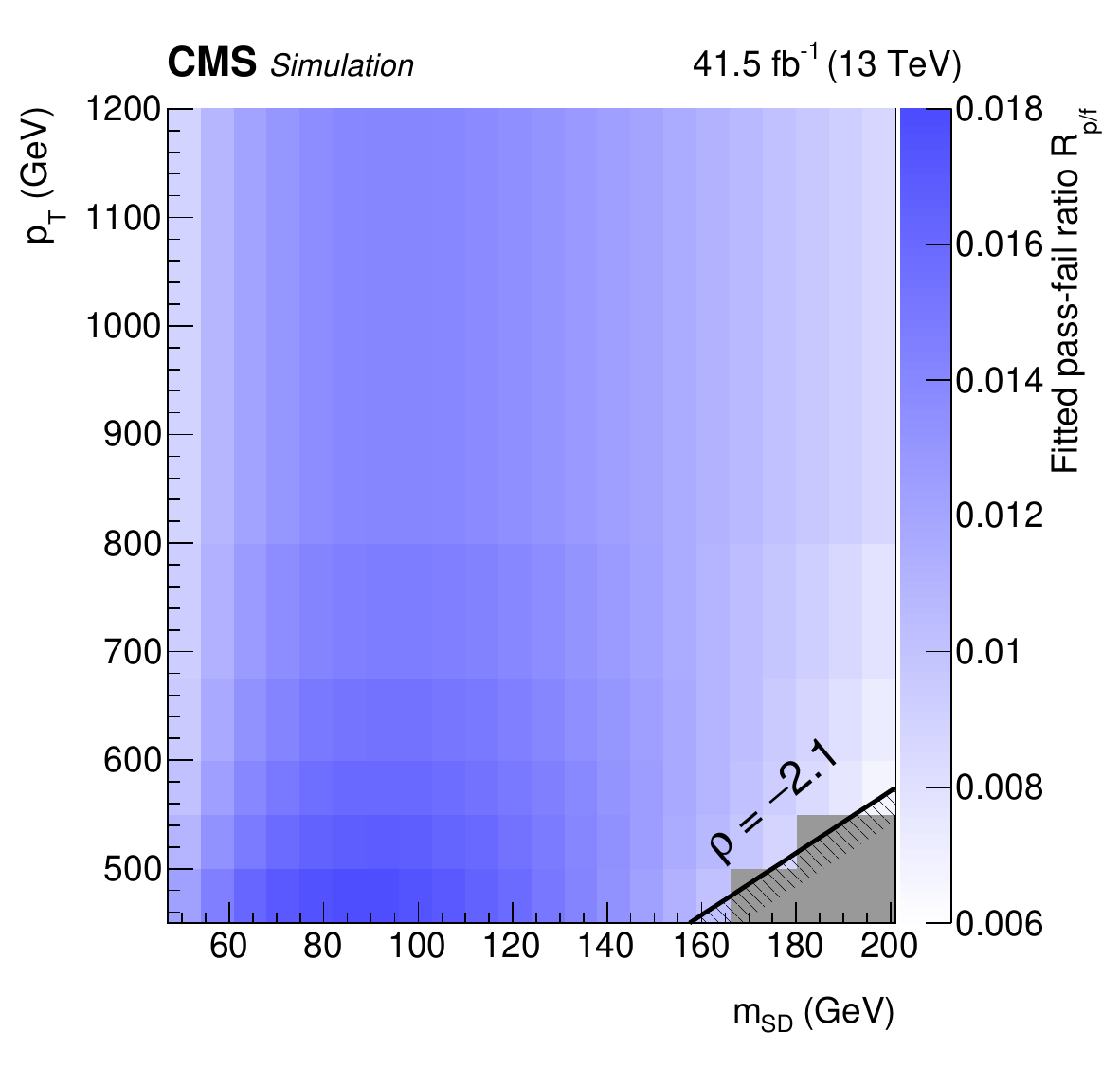}
  \caption{The fitted pass-fail ratio \Rpf as a function of jet \pt and \mSD for data collected in 2017.
  The ratio relates the QCD multijet event yield in the DDBT passing region to that of the failing region.
  The binning corresponds to the 22 \mSD bins and 6 \pt categories used in the statistical analysis.
  The lower-right bins filled in gray fall outside of the $\rho$ acceptance.
  }
  \label{fig:dataRatio2d}
\end{figure*}

The 2017 fitted pass-fail ratio \Rpf as a function of \mSD and \pt under the signal-plus-background hypothesis is shown in Fig.~\ref{fig:dataRatio2d}.
In the absence of correlations between \mSD, \pt, and the DDBT efficiency, the ratio would be approximately 0.01.
The majority of the difference from 0.01 is a result of the expected pass-fail ratio, which ranges from 0.007 to 0.018, while the data residual correction ranges from 0.86 to 1.05.
The other data taking periods are similar.
As discussed in Section~\ref{sec:unc}, the components of the pass-fail ratio are among the largest sources of uncertainty in the analysis.

As the QCD background estimate relies solely on the properties of the $\PH(\bbbar)$ candidate jet, \Vjets proceses in which the candidate jet does not arise from a vector boson decay are included in this estimate, and therefore are removed from the predicted yields of those processes.

In order to validate the background estimation method and associated systematic uncertainties, bias studies are performed using an alternative functional form for the pass-fail ratio in the background model.
Pseudo-experiment data sets are generated assuming the alternative background model, with the injection of signal events for a range of hypothetical signal strength values of between 0 and 5 times the SM expectation, and then fit with the nominal signal-plus-background model.
No significant bias in the fitted signal strength is observed; specifically, the means of the differences between the fitted and injected signal strengths divided by the fitted uncertainty are found to be less than 15\%.
Therefore, no additional systematic uncertainty is assigned for this potential bias from the background modeling.

\section{Systematic uncertainties}
\label{sec:unc}
The systematic uncertainties associated with the jet mass scale, the jet mass resolution, and the $\nddt$ selection efficiency are correlated among the \PW, \PZ, and $\PH(\bbbar)$ processes.
These uncertainties are estimated in data using an independent sample of merged \PW boson jets in semileptonic \ttbar events, where the hadronically decaying \PW boson is reconstructed as a single AK8 jet.

For this sample, data events are required to have an energetic muon with $\pt > 100\GeV$ and $\abs{\eta} < 2.1$ , $\ptmiss > 80\GeV$, a high-\pt AK8 jet with $\pt > 200\GeV$, and an additional \cPqb-tagged AK4 jet separated from the AK8 jet by $\Delta R > 0.8$ with $\pt>30\GeV$.
Using the same $\nddt$ requirement applied in the signal regions, we define two samples, one with events that pass and one with events that fail the $\nddt$ selection, for merged \PW boson jets in data and simulation.
A simultaneous fit to the two samples in \mSD is performed in order to extract the selection efficiency of a merged \PW jet in simulation and in data.
The data-to-simulation scale factors for the $\nddt$ selection efficiency are measured separately for the three data taking periods, as listed in Table~\ref{tab:SF}.

The jet mass scale and jet mass resolution data-to-simulation scale factors are extracted from the same fit, and are also shown in Table~\ref{tab:SF}.
As the semileptonic \ttbar sample does not contain a large population of very energetic jets, an additional systematic uncertainty is included to account for the extrapolation to very high \pt jets.
This additional uncertainty is estimated to be $0.5\%$ per $100\GeV$, based on a study of fitting the \mSD distributions of merged top quark jets in different \pt ranges above 350\GeV~\cite{Sirunyan:2019vxa}.
In total, the jet mass scale uncertainty increases with jet \pt, ranging from $1.2\%$ at 450\GeV to $2.1\%$ at 800\GeV.
While the jet mass scale and resolution among the different years of data collection are similar, their data-to-simulation scale factors and uncertainties vary because of the different generator tunes used in the simulations.

The uncertainty on the efficiency of the DDBT is estimated using data and simulation samples enriched in $\bbbar$ pairs from gluon splitting~\cite{Sirunyan:2017ezt}.
The gluon splitting samples require that both subjets of an AK8 jet contain a muon, targeting semileptonic decays of the \PQb hadrons.
The method is based on yields extracted from fits to the distributions of the jet probability tagger~\cite{Chatrchyan:2012jua,Sirunyan:2017ezt} discriminant, which uses the signed impact parameter significance of the tracks associated with the jet to obtain a likelihood for the jet to originate from the primary vertex.

Given that the DDBT efficiencies could differ between $\bbbar$ jets from gluon splitting and from color-singlet \PZ or Higgs boson decays, the efficiencies extracted from the gluon splitting samples are used only to estimate the uncertainty on the DDBT efficiency, and are not used to correct the efficiency.
The applied DDBT data-to-simulation scale factor is included in the signal extraction fit as a constrained nuisance parameter, with a nominal value of unity and an uncertainty equal to the difference between the DDBT data-to-simulation scale factor and unity, as shown in Table~\ref{tab:SF}.
The scale factor is further constrained via the observed \PZ boson yield in the passing and failing regions.
This strategy differs from that of the previous CMS analysis~\cite{Sirunyan:2017dgc}, resulting in an increase in the post-fit systematic uncertainty of the tagger efficiency from 4\% to about 14\%.

\begin{table*}[htb]
  \centering
  \topcaption{Summary of applied data-to-simulation scale factors for the jet mass scale, jet mass resolution, $\nddt$ selection, and DDBT selection for different data taking periods.}
\cmsTable{
  \begin{tabular}{cccccc}
    Data & Integrated & \multirow{ 2}{*}{Jet mass scale} & \multirow{ 2}{*}{Jet mass resolution} & \multirow{ 2}{*}{$\nddt$ selection} & DDBT selection \\
    period & luminosity (\fbinv) & & & & ($\Pg\to\bbbar$)    \\    \hline
    2016  & 35.9 & $ 1.000 \pm 0.012$ & $1.084 \pm 0.091$ &  $ 0.993 \pm 0.043$ & $1.00 \pm 0.23$ \\
    2017  & 41.5 & $ 0.987 \pm 0.012$ & $0.905 \pm 0.048$ &  $ 0.924 \pm 0.018$ & $1.00 \pm 0.32$ \\
    2018  & 59.2 & $ 0.970 \pm 0.012$ & $0.908 \pm 0.014$ &  $ 0.953 \pm 0.016$ & $1.00 \pm 0.30$ \\
  \end{tabular}}
  \label{tab:SF}
\end{table*}

The scale factors described above determine the initial distributions of the jet mass for the $\PW(\qqbar)$, $\PZ(\qqbar)$, and $\PH(\bbbar)$ processes.
In the fit to data, the jet mass scales and resolutions are treated as constrained nuisance parameters with nominal values and uncertainties as shown in Table~\ref{tab:SF}, and are further constrained by the presence of the $\PV$ resonances in the jet mass distribution.
A single nuisance parameter per year is considered for the $\nddt$ selection efficiency uncertainty.
Alternative configurations in which multiple nuisance parameters are considered for the $\nddt$ selection efficiency  uncertainty in order to account for a potential mass or \pt dependence were found to have no impact on the analysis results.

The uncertainty associated with the choice of QCD renormalization and factorization scales in the modeling of \ggH production is propagated to the total expected yield of the \ggH signal via varying each factor by one-half or two around the nominal value and finding the envelope of all combinations of such variations, except those where one scale is multiplied by 0.5 and the other is multiplied by 2~\cite{Cacciari:2003fi,Catani:2003zt}.
This results in a 30\% uncertainty for the \POWHEG sample with \pt reweighting~\cite{Sirunyan:2017dgc} and a 20\% uncertainty for the \HJMINLO sample.
These variations account for the effect on both the inclusive cross section and acceptance.
An additional uncertainty is considered for the reweighted \POWHEG sample, in which the shape of the \ggH Higgs boson \pt distribution is allowed to vary by a linear function of the Higgs boson \pt that changes the relative yield at 1.2\TeV by $\pm$30\% for a $1\,\sigma$ effect, without changing the overall yield.
Uncertainties related to finite top quark mass effects are estimated in Ref.~\cite{Lindert:2018iug}, and are found to be subdominant to the scale uncertainties for the \HJMINLO sample.
For the $\PV(\qqbar)$ yield, two nuisance parameters account for potential \pt-dependent deviations due to missing higher-order corrections, where one is 10\% in magnitude on the total yield, and the other increases from 0 to 7\% versus \pt~\cite{Sirunyan:2017jix,Denner:2009gj,Denner:2011vu,Denner:2012ts,Kuhn:2005gv,Kallweit:2014xda,Kallweit:2015dum}.
An additional systematic uncertainty of 2 to 6\%, depending on \pt, is included to account for potential differences between the higher-order corrections to the \PW and \PZ cross sections (EW $\PW/\PZ$ decorrelation)~\cite{Sirunyan:2017jix}.

Finally, systematic uncertainties are applied to the $\PW(\qqbar)$, $\PZ(\qqbar)$, \ttbar, and $\PH(\bbbar)$ yields to account for the uncertainties due to the jet energy scale and resolution~\cite{jec} and the limited simulation sample sizes.
The effect of limited QCD simulation sample size on the separate fit to determine the expected pass-fail ratio $\epsilon^{\mathrm{QCD}}(\rho,\,\pt)$ is also included.
Other experimental uncertainties, including those related to the determination of the integrated luminosity~\cite{lumi}, variations in the amount of pileup, modeling of the trigger acceptance, and the isolation and identification of leptons are also considered.
Table~\ref{tab:systematics} lists the major sources of uncertainty and their observed impact on the Higgs boson signal strength $\mu_{\PH}$, defined as the ratio of the measured to the SM expected $\PH(\bbbar)$ production, in the combined fit.
One of the largest sources of statistical uncertainty is the data residual correction to the pass-fail ratio $\Rpf$, while the largest source of systematic uncertainty is the expected pass-fail ratio $\epsilon^{\mathrm{QCD}}$, which is initially estimated from simulation and further constrained by the data.
Overall, the $\mu_{\PH}$ measurement is limited by statistical sources of uncertainty.

\begin{table*}[htb]
\centering
\topcaption{Major sources of uncertainty in the measurement of the signal strength $\mu_{\PH}$ based on the \HJMINLO prediction, and their observed impact ($\Delta\mu_{\PH}$) from a fit to the combined data set.
Decompositions of the statistical, systematic, and theoretical components of the total uncertainty are specified.
The impact of each uncertainty is evaluated by computing the uncertainty excluding that source and subtracting it in quadrature from the total uncertainty.
The sum in quadrature for each source does not in general equal the total uncertainty of each component because of correlations in the combined fit between nuisance parameters corresponding to different sources.}
\begin{tabular}{ l l l }
Uncertainty source & \multicolumn{2}{c}{$\Delta\mu_{\PH}$} \\
\hline
  Statistical                                         & $+\muStatErrMINLO$ & $-\muStatErrMINLO$ \\
  \qquad Signal extraction                            & $+0.9$ & $-0.8$\\
  \qquad QCD pass-fail ratio (data correction)        & $+0.8$ & $-0.7$\\
  \qquad \ttbar normalization and misidentification   & $+0.4$ & $-0.4$ \\[\cmsTabSkip]
  Systematic                                          & $+\muSystErrHiMINLO$ & $-\muSystErrLoMINLO$\\
  \qquad QCD pass-fail ratio (simulation)             & $+0.6$ & $-0.6$\\
  \qquad DDBT efficiency                              & $+0.3$ & $-0.1$\\
  \qquad Jet mass scale and resolution                & $+0.3$ & $-0.3$\\
  \qquad Jet energy scale and resolution              & $+0.1$ & $-0.1$\\
  \qquad Simulated sample size                        & $+0.2$ & $-0.1$\\
  \qquad Other experimental uncertainties             & $+0.1$ & $-0.1$\\[\cmsTabSkip]
  Theoretical                                         & $+\muThyErrHiMINLO$ & $-\muThyErrLoMINLO$\\
  \qquad \Vjets modeling                              & $+0.6$ & $-0.4$\\
  \qquad \PH modeling                                 & $+0.5$ & $-0.3$\\[\cmsTabSkip]
  Total                                               & $+\muErrHiMINLO$ & $-\muErrLoMINLO$\\
\end{tabular}
\label{tab:systematics}
\end{table*}

\section{Results}
\label{sec:results}
A binned maximum likelihood fit to the observed \mSD distributions is performed using the sum of the signal and background contributions.
The fit is performed simultaneously in the DDBT passing and failing regions of the six \pt categories, as well as in the DDBT passing and failing components of the \ttbar-enriched control region.
The fit is performed separately for the three year periods.
A combined fit over the three periods is performed for the final result.
The theoretical uncertainties are correlated between different years.
The test statistic chosen to determine the signal yield is based on the profile likelihood ratio~\cite{LHCCLs}.
Systematic uncertainties are incorporated into the analysis via nuisance parameters and treated according to the frequentist paradigm.
The best-fit value of each signal strength parameter and an approximate 68\% confidence level (\CL) interval are extracted following the procedure described in Section~3.2 of Ref.~\cite{Khachatryan:2014jba}.

\begin{figure*}[hbtp]
\centering
    \includegraphics[width=\cmsSmallFigWidth]{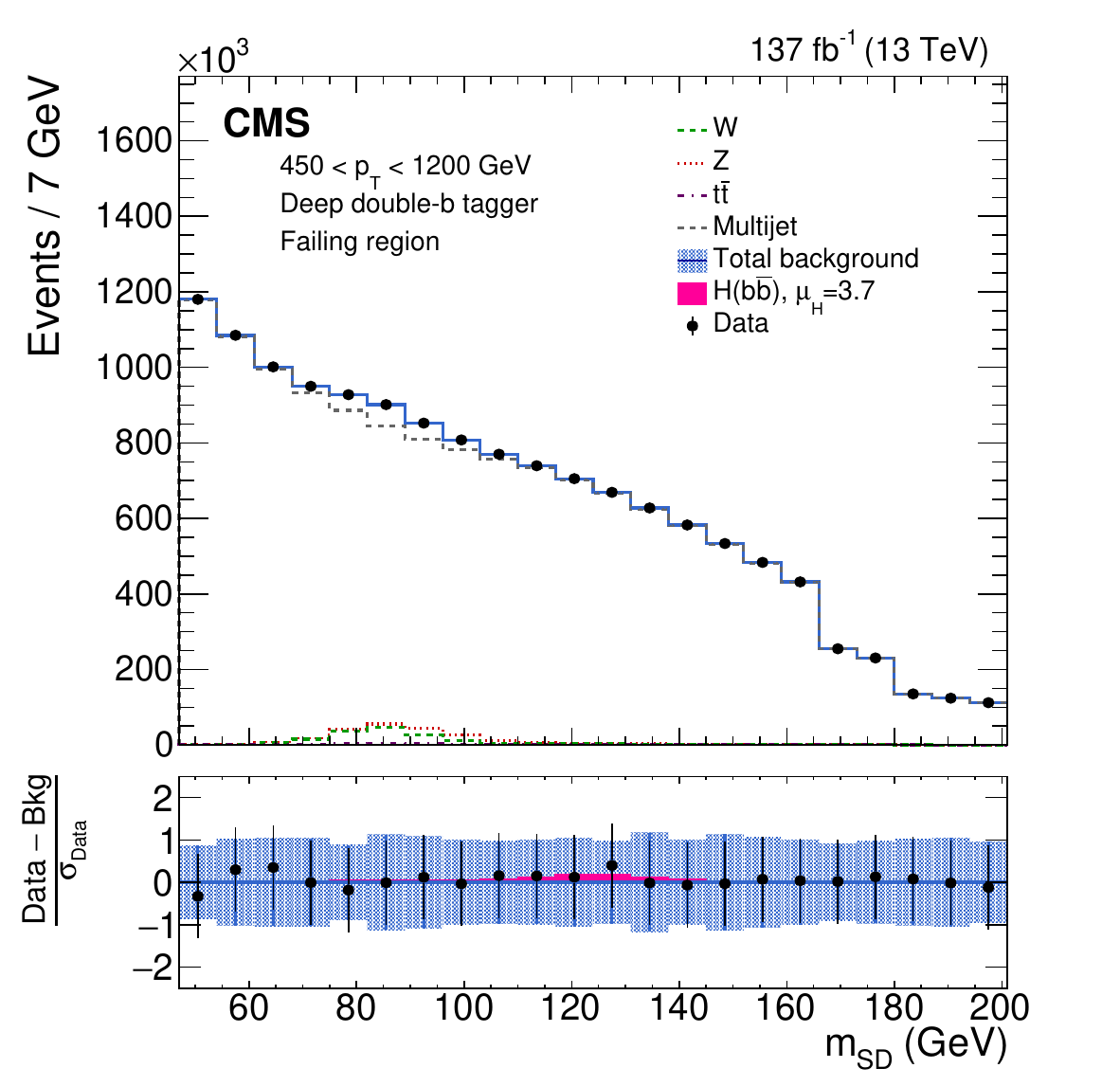}
    \includegraphics[width=\cmsSmallFigWidth]{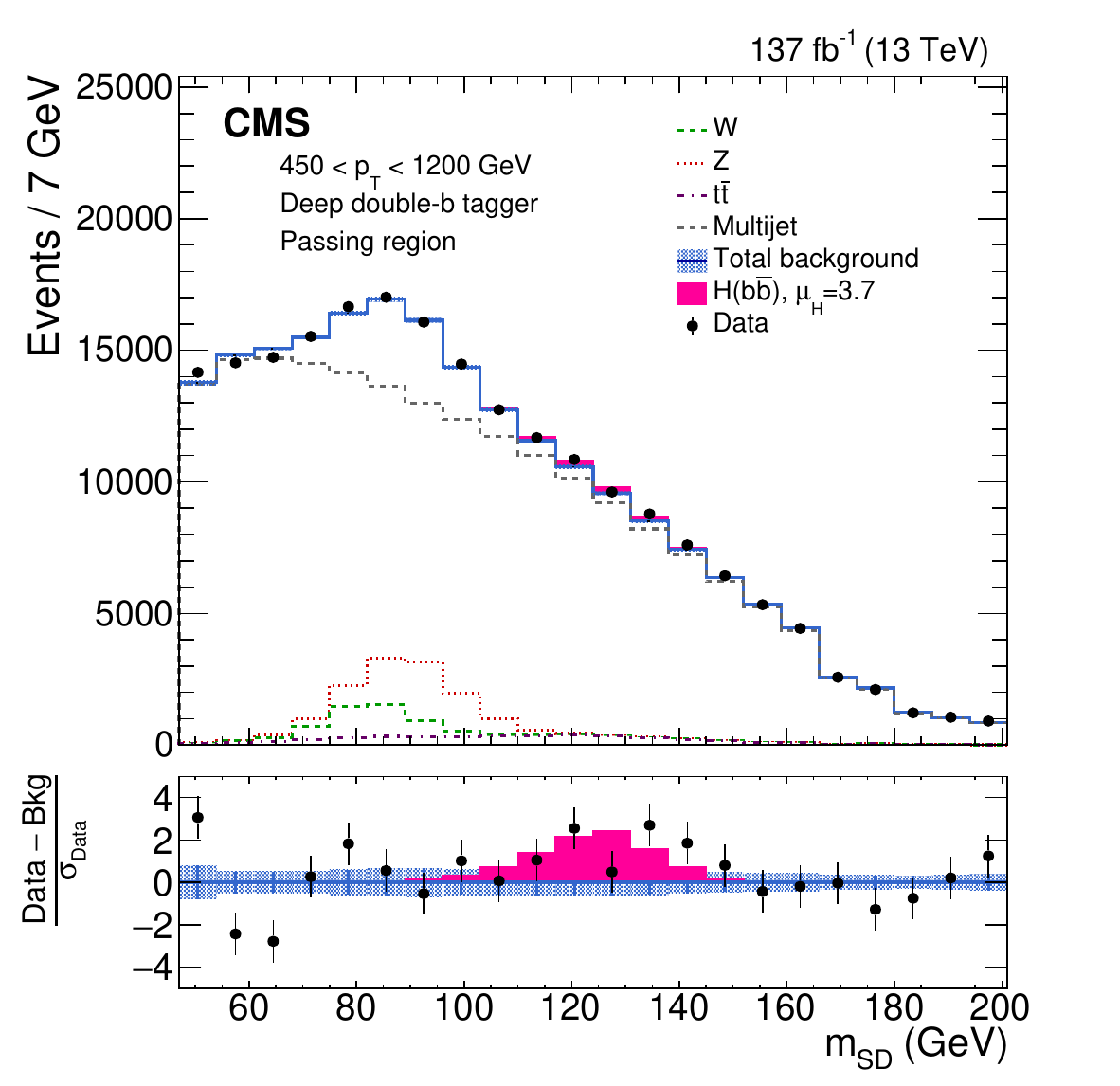}
    \caption{
      The observed and fitted background \mSD distributions for the DDBT failing (left) and passing (right) regions, combining all the \pt categories, and three data collection years.
      The fit is performed under the signal-plus-background hypothesis with one inclusive $\PH(\bbbar)$ signal strength parameter floating in all the \pt categories.
      Because of the finite $\rho$ acceptance, some \mSD bins within a given \pt category may be removed, giving rise to the steps at 166 and 180\GeV.
      The shaded blue band shows the systematic uncertainty in the total background prediction.
      The bottom panel shows the difference between the data and the total background prediction, divided by the statistical uncertainty in the data.
      In the failing region, the background model includes a free parameter for each \mSD bin, ensuring the nearly perfect agreement between the model and the data---this agreement is imperfect because the passing region is fit simultaneously and the global best fit is a balance between the two regions.
      Thus, the statistical uncertainty in the data gives rise to the systematic uncertainty in the background prediction.
      This is reflected in the fact that the error bar for the data and the uncertainty band for the background are approximately equal in size.
    }
 \label{fig:results}
 \end{figure*}

\begin{figure*}[hbtp]
  \centering
  \includegraphics[width=\cmsSmallFigWidth]{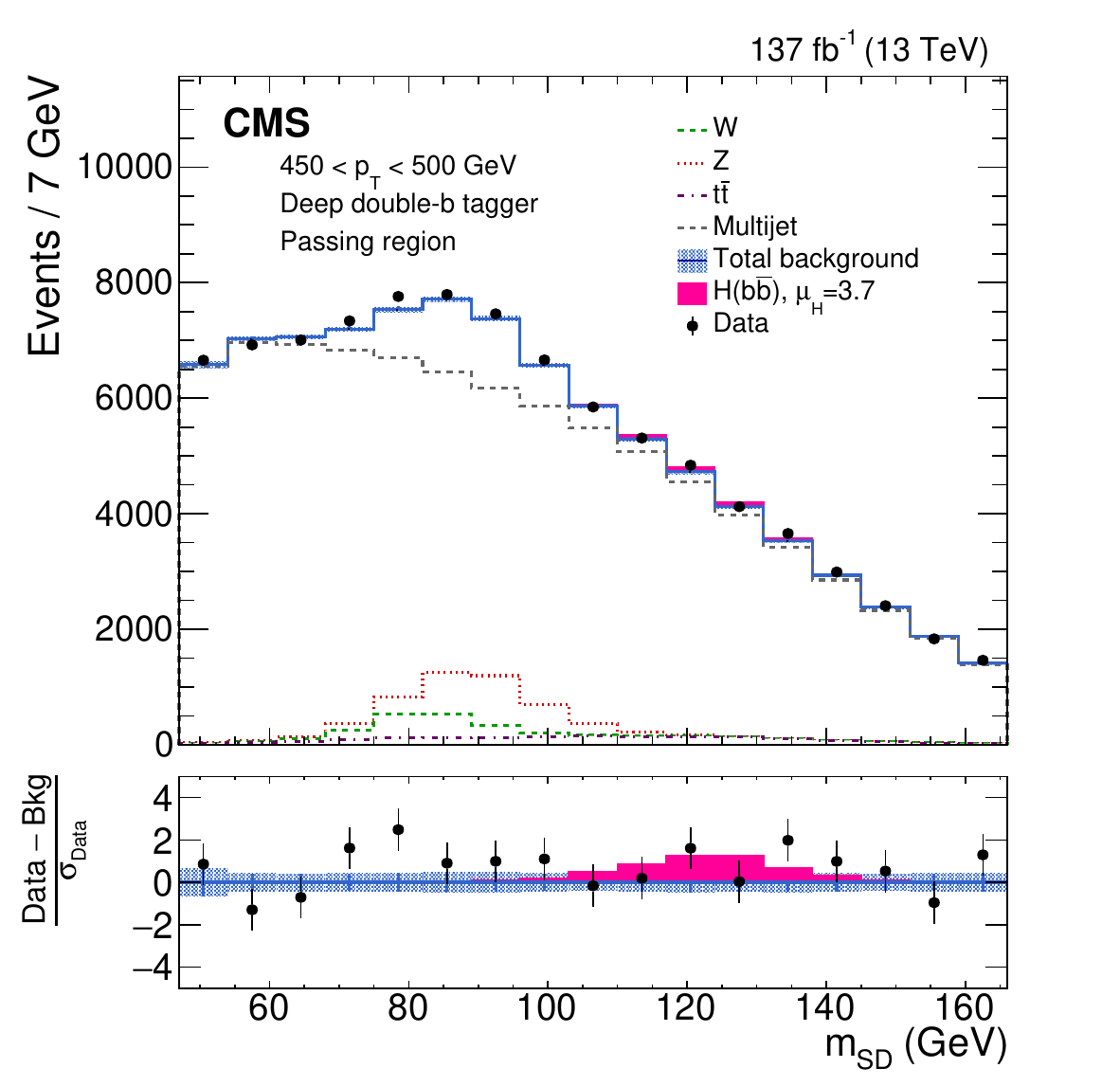}
  \includegraphics[width=\cmsSmallFigWidth]{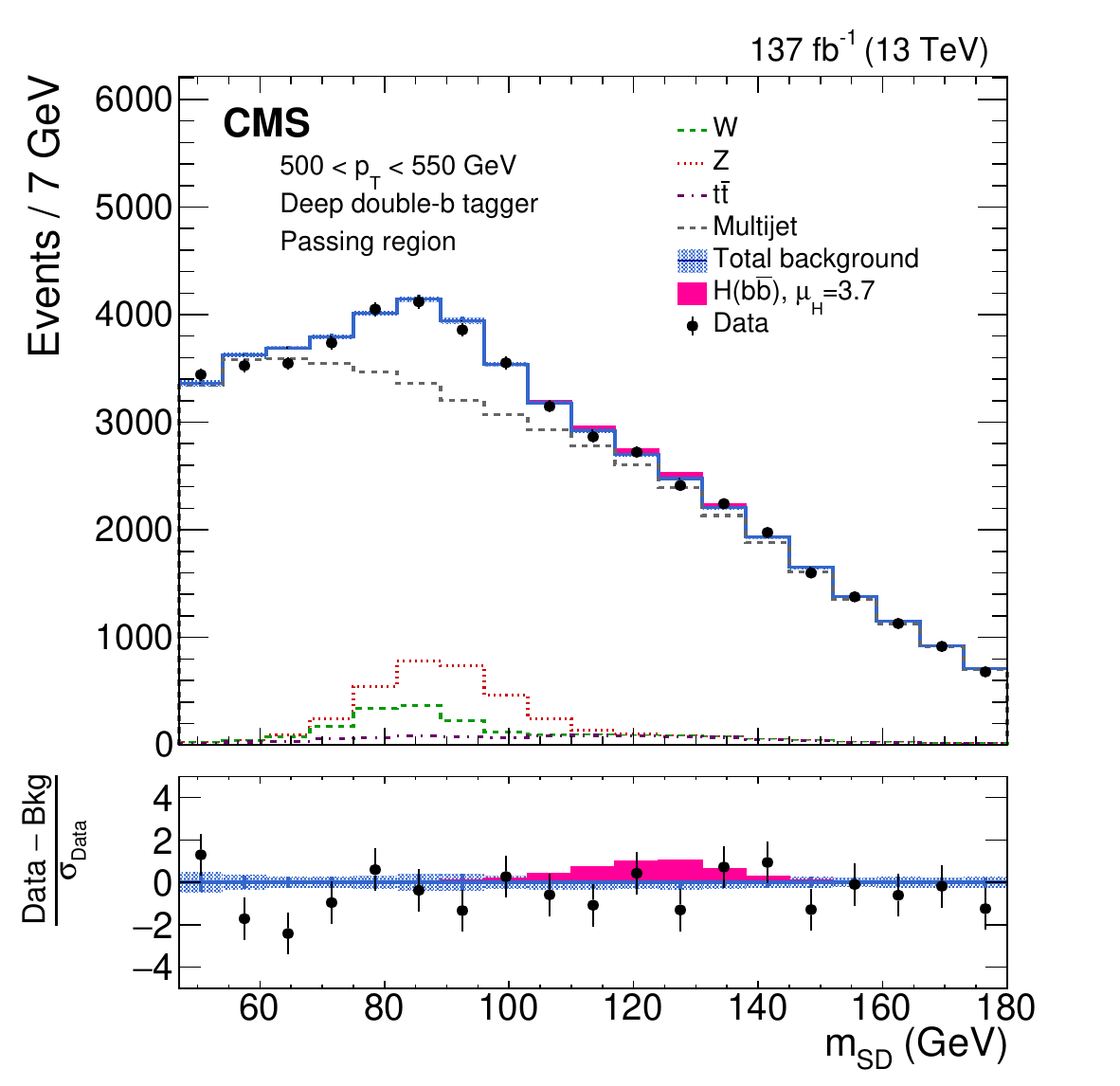}\\
  \includegraphics[width=\cmsSmallFigWidth]{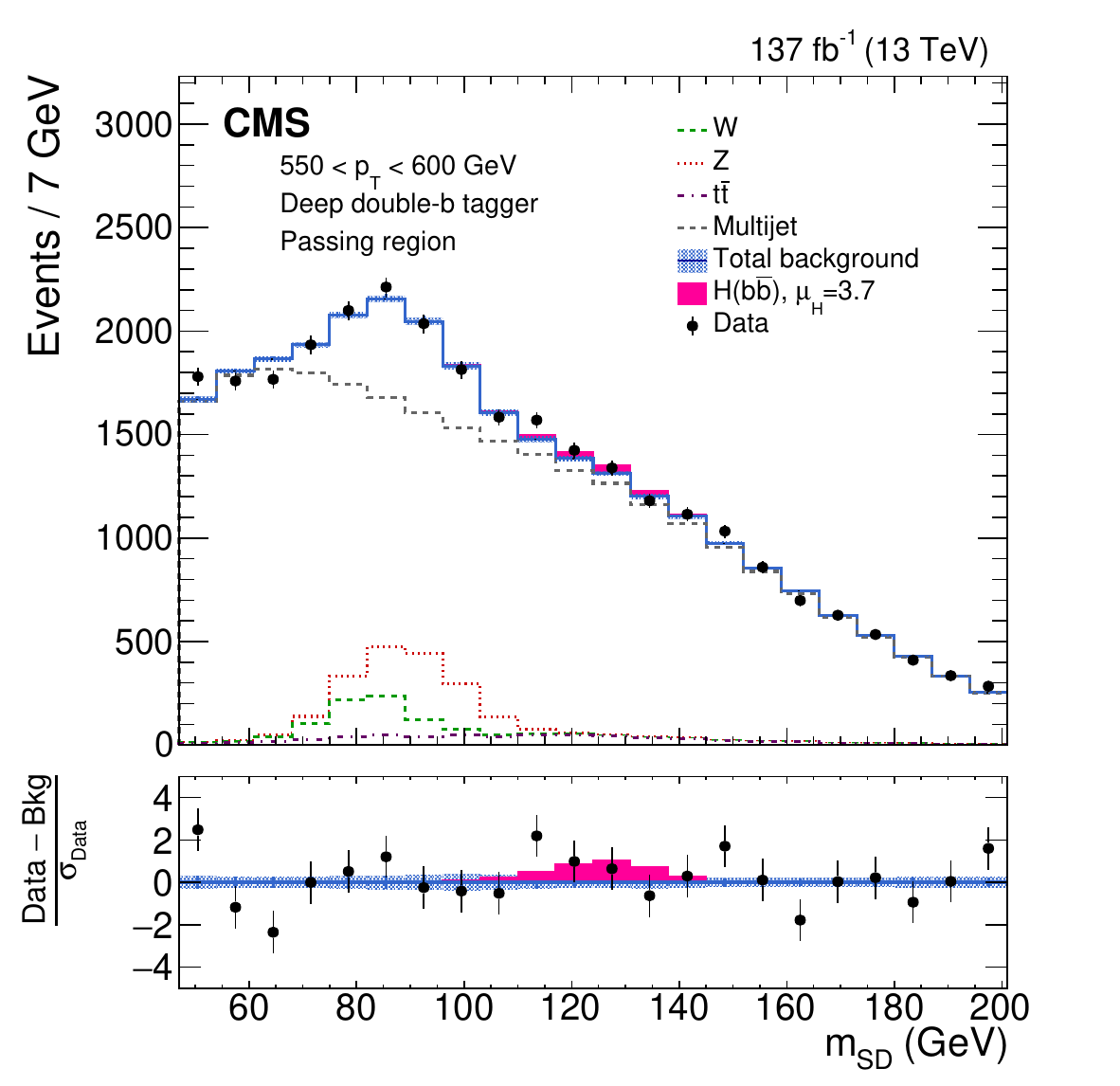}
  \includegraphics[width=\cmsSmallFigWidth]{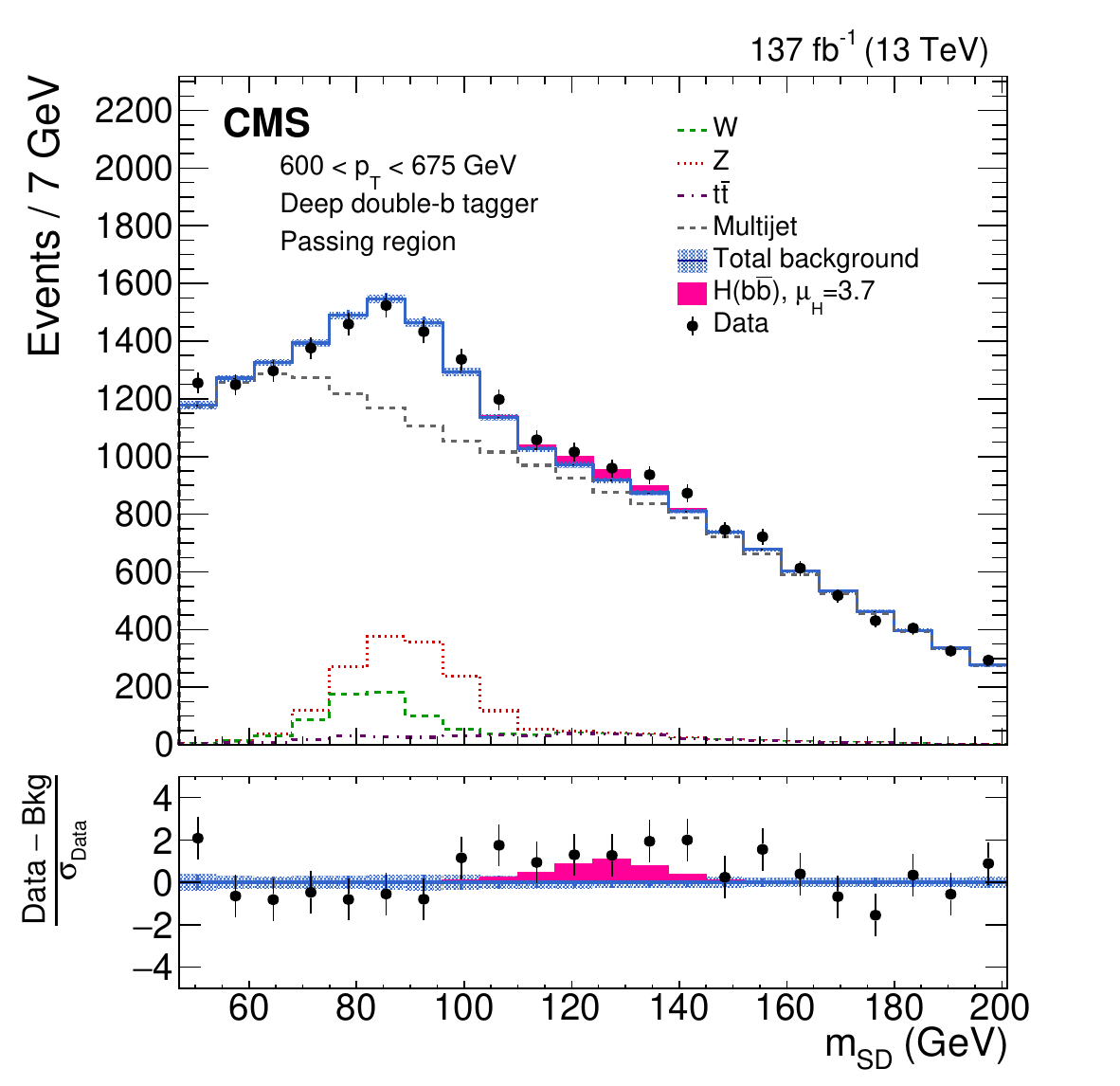}\\
  \includegraphics[width=\cmsSmallFigWidth]{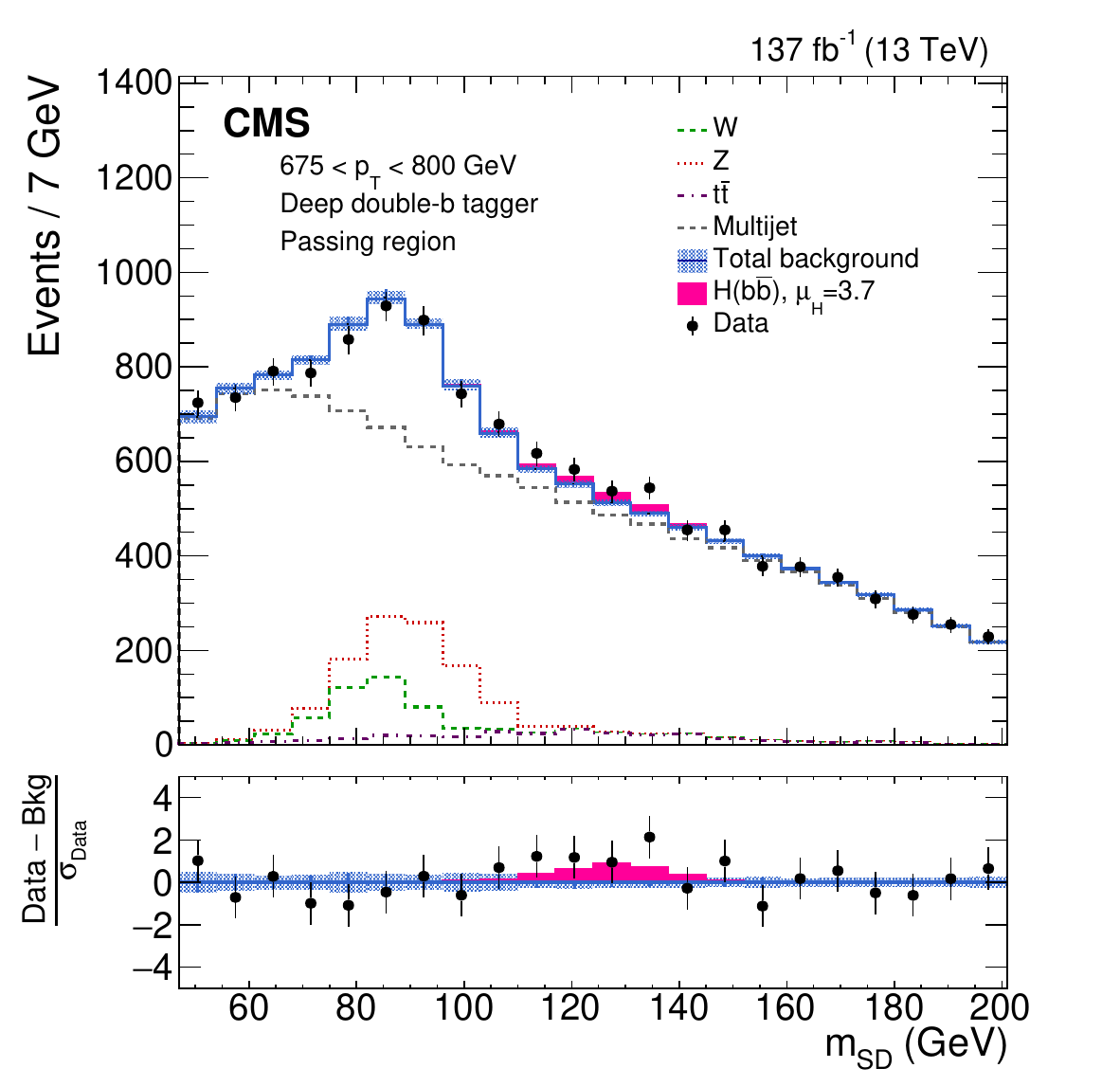}
  \includegraphics[width=\cmsSmallFigWidth]{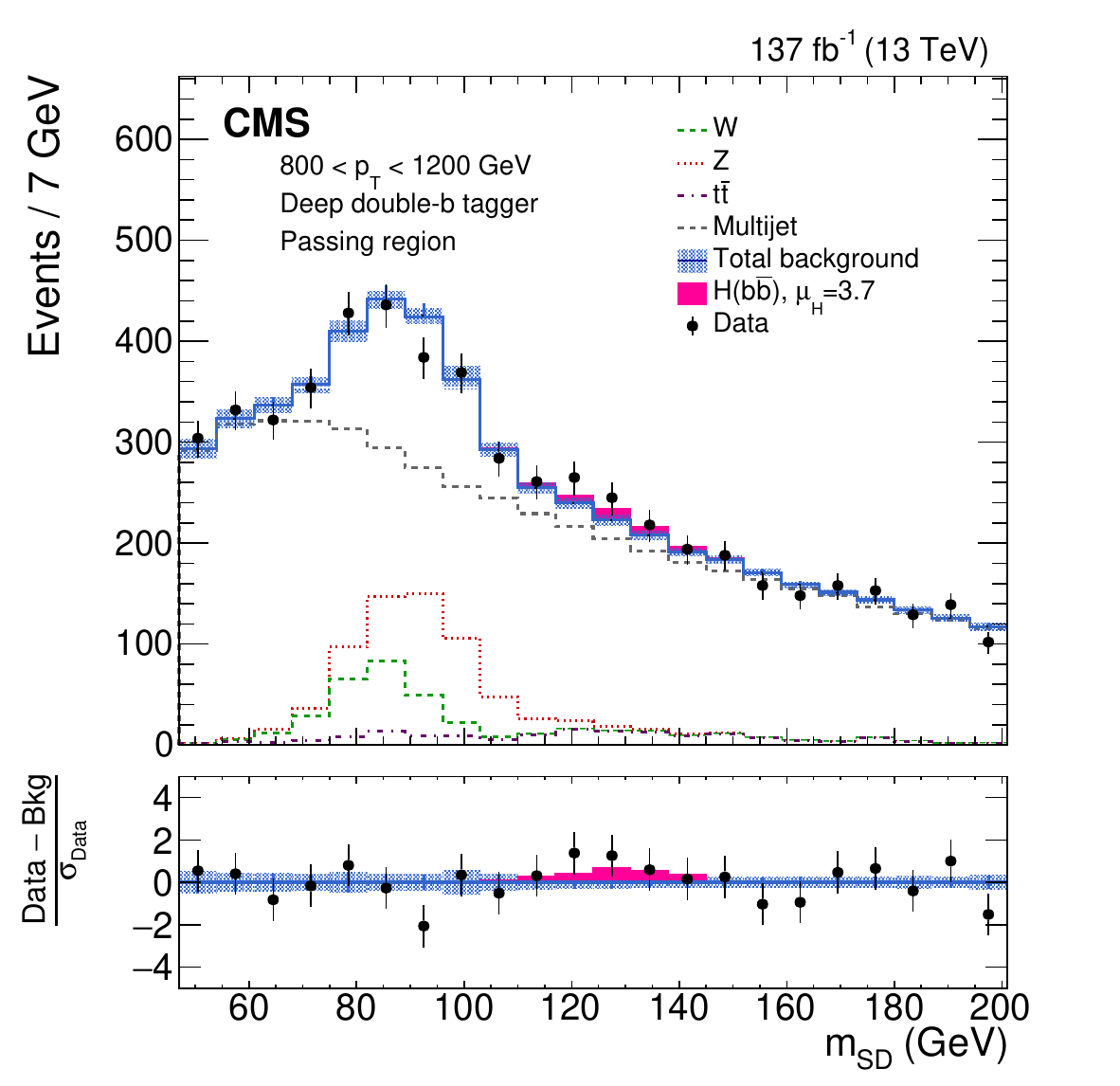}\\
  \caption{
    The observed and fitted background \mSD distributions in each
    \pt category in the DDBT passing regions.
    The fit is performed under the signal-plus-background hypothesis with one inclusive $\PH(\bbbar)$ signal strength parameter floating in all the \pt categories.
    The shaded blue band shows the systematic uncertainty in the total background prediction.
    The bottom panel shows the difference between the data and the total background prediction, divided by the statistical uncertainty in the data.
  }
  \label{fig:results_pt}
\end{figure*}

Figure~\ref{fig:results} shows the \mSD distributions in the combined data set for the DDBT passing and failing regions with the fitted background.
The bottom panels of Fig.~\ref{fig:results} show the difference between the data and the prediction from the background, divided by the statistical uncertainty in the data.
These highlight the contributions from Higgs and \PV boson production in the failing and passing regions.
The \PW boson contribution in the passing region is due to the misidentification of \PW(\qqbar) decays by the DDBT.
The agreement between the data and the signal-plus-background model is quantified with a Kolmogorov-Smirnov goodness-of-fit test~\cite{KS}, which yields a p-value of 17\%.
In Fig.~\ref{fig:results_pt}, the \mSD distributions are shown for each \pt category in the passing region.
The nuisance parameters related to the jet mass scale uncertainties, whose values extend up to 2\GeV in the case of the \PZ boson as discussed in Section~\ref{sec:unc}, do not significantly deviate from their pre-fit expectations.

To validate the substructure and \PQb tagging techniques employed in this search, a maximum likelihood fit is performed using a model where the \PZ(\qqbar) signal strength ($\mu_{\PZ}$) and $\mu_{\PH}$ are left unconstrained.
In the DDBT passing region, decays of the \PZ boson to $\bbbar$ constitute 79\% of all \PZ decays.
The product of cross section and branching fraction for the $\PZ(\qqbar)$ sample with \pt of the $\PZ$ boson greater than $300\GeV$ is 15.9\unit{pb} and the product of acceptance and efficiency for events in which the $\PZ$ boson is matched to the $\PH(\bbbar)$ candidate jet in the DDBT passing region is 0.41\%.
The measured $\mu_{\PZ}$ value is $\muZVal\pm\muZStatErr\stat_{-\muZSystErrLo}^{+\muZSystErrHi}\syst_{-\muZThyErrLo}^{+\muZThyErrHi}\thy$.
This demonstrates that the \PZ boson is clearly separable from the background.
In this measurement, the dominant source of systematic uncertainty is the DDBT scale factor.
For the remainder of results, $\mu_{\PZ}$ is fixed to its expectation, with the corresponding uncertainties, as described in Section~\ref{sec:unc}.
Thus, the \PZ boson resonance is used to further constrain the DDBT scale factor in the Higgs boson measurements.

To extract the Higgs boson signal, three maximum likelihood fits are performed to the data, each with a different degree of reliance on the modeling of the Higgs boson \pt spectrum:
the nominal inclusive fit using one $\mu_{\PH}$ parameter for all \PH production modes and all jet \pt categories, an alternative fit using an independent $\mu_{\PH}$ parameter for each \pt category for all \PH production modes to assess the compatibility among the \pt categories, and a fit which unfolds detector effects to present results for the \ggH production mode at the generator level.

{\tolerance=1000
The product of cross section and branching fraction for all $\PH(\bbbar)$ processes with Higgs boson $\pt>300\GeV$ is 0.12\unit{pb} and the product of acceptance and efficiency for events in which the $\PH$ boson is matched to the $\PH(\bbbar)$ candidate jet in the DDBT passing region is 1.7\%.
In the inclusive fit using the \HJMINLO sample as the \ggH signal model and including the contributions from the other production modes, the measured $\mu_{\PH}$ value is $\muValMINLO\pm\muStatErrMINLO\stat_{-\muSystErrLoMINLO}^{+\muSystErrHiMINLO}\syst_{-\muThyErrLoMINLO}^{+\muThyErrHiMINLO}\thy$.
Upper limits at 95\% \CL using the \CLs criterion~\cite{CLS1,CLS2} are obtained using asymptotic formulae~\cite{Cowan:2010js}.
The corresponding observed and expected upper limits on $\mu_{\PH}$ at a 95\% \CL are \muObsLimitMINLO and \muExpLimitMINLO, respectively, while the observed and expected significances~\cite{pvalue} with respect to the background-only hypothesis are $\muObsSigMINLO\,\sigma$ and $\muExpSigMINLO\,\sigma$.
The measurement exhibits an excess over the SM expectation ($\mu_{\PH}=1$), with a significance of $\muObsSigMINLOMuOne\,\sigma$.
Table~\ref{tab:ObservedSig} summarizes the measured signal strengths and significances for the Higgs and \PZ boson processes.
The primary results using the \ggH \pt spectrum from \HJMINLO~\cite{Hamilton:2012rf,Becker:2669113} are shown, alongside results using the \ggH \pt spectrum from Ref.~\cite{Sirunyan:2017dgc} for ease of comparison.
The prediction used for the \ggH \pt spectrum in Ref.~\cite{Sirunyan:2017dgc} is different from that of \HJMINLO in both shape and total cross section, which is primarily due to the different accuracy of finite top quark mass correction included in the simulation.
In particular, the number of \ggH signal events predicted by \HJMINLO in the fiducial region of the analysis is approximately a factor of two smaller than that of Ref.~\cite{Sirunyan:2017dgc}, which is reflected in the fitted $\mu_{\PH}$ values and their uncertainties.
The fitted signal strength value and its uncertainty are sensitive to the \ggH theoretical prediction and associated uncertainty, which are challenging to obtain in the high-\pt regime.
\par}

\begin{table}[hbtp]
\centering
\topcaption{
    Fitted signal strength, and expected and observed significance of the Higgs and \PZ boson signals.
    The Higgs boson results are presented with two \ggH signal models, one using the nominal \HJMINLO sample and the other simulated with the same procedure described in Ref.~\cite{Sirunyan:2017dgc}.
    The 95\% confidence level upper limit (UL) on the Higgs boson signal strength is also listed.
    In the results for the Higgs boson, the \PZ boson yield is fixed to the SM prediction value with the corresponding theoretical uncertainties to better constrain the data-to-simulation scale factor for the DDBT.
    For the expected and observed signal strengths of the \PZ boson, the Higgs boson signal strength is freely floating.
}
\begin{tabular}{ lllll }
                                                  & 2016  & 2017                   & 2018                        & Combined \\
\hline
Expected $\mu_{\PZ}$               &  $1.00_{-0.28}^{+0.38}$ & $1.00_{-0.29}^{+0.42}$ & $1.00_{-0.29}^{+0.43}$      & $1.00_{-0.19}^{+0.23}$\\
Observed $\mu_{\PZ}$               &  $0.86_{-0.24}^{+0.32}$ & $1.11_{-0.33}^{+0.48}$ & $0.91_{-0.26}^{+0.37}$      & $\muZVal_{-\muZErrLo}^{+\muZErrHi}$\\ [\cmsTabSkip]
\HJMINLO~\cite{Hamilton:2012rf,Becker:2669113}                                     &                         &                        &                             &\\
Expected $\mu_{\PH}$               &  $1.0_{-3.5}^{+3.3}$ & $1.0\pm 2.5$  & $1.0_{-2.4}^{+2.3}$    & $1.0\pm 1.4$\\
Observed $\mu_{\PH}$               &  $7.9_{-3.2}^{+3.4}$ & $4.8_{-2.5}^{+2.6}$  & $1.7\pm 2.3$    & $\muValMINLO_{-\muErrLoMINLO}^{+\muErrHiMINLO}$\\
Expected \PH significance ($\mu_{\PH}=1$)        &  $0.3\,\sigma$           & $0.4\,\sigma$            & $0.4\,\sigma$              & $\muExpSigMINLO\,\sigma$          \\
Observed \PH significance                        &  $2.4\,\sigma$           & $1.9\,\sigma$            & $0.7\,\sigma$              & $\muObsSigMINLO\,\sigma$          \\
Expected UL $\mu_\PH$ ($\mu_{\PH}=0$)        &  $<$6.8                & $<$5.0                 & $<$4.7                   & $<$\muExpLimitMINLO   \\
Observed UL $\mu_\PH$                        &  $<$13.9                & $<$9.3                 & $<$5.9                   & $<$\muObsLimitMINLO   \\[\cmsTabSkip]
Ref.~\cite{Sirunyan:2017dgc} \PH \pt spectrum                   &                         &                        &                             &\\
Expected $\mu_{\PH}$               &  $1.0\pm 1.5$ & $1.0_{-1.0}^{+1.1}$ & $1.0_{-1.0}^{+1.1}$    & $1.0_{-0.6}^{+0.7}$  \\
Observed $\mu_{\PH}$               &  $4.0_{-1.6}^{+1.9}$ & $2.2_{-1.2}^{+1.4}$ & $1.1\pm 1.1$    & $\muVal_{-\muErrLo}^{+\muErrHi}$  \\
Expected \PH significance ($\mu_{\PH}=1$)        &  $0.7\,\sigma$           & $0.9\,\sigma$           & $1.0\,\sigma$              & $\muExpSig\,\sigma$          \\
Observed \PH significance                        &  $2.6\,\sigma$           & $1.8\,\sigma$           & $1.1\,\sigma$              & $\muObsSig\,\sigma$          \\
Expected UL $\mu_\PH$ ($\mu_{\PH}=0$)        &  $<$3.4                & $<$2.4                & $<$2.3                   & $<$\muExpLimit               \\
Observed UL $\mu_\PH$                        &  $<$7.4                & $<$4.6                & $<$3.2                   & $<$\muObsLimit               \\
\end{tabular}
\label{tab:ObservedSig}
\end{table}

\begin{figure*}[hbtp]
\centering
\includegraphics[width=\cmsSmallFigWidth]{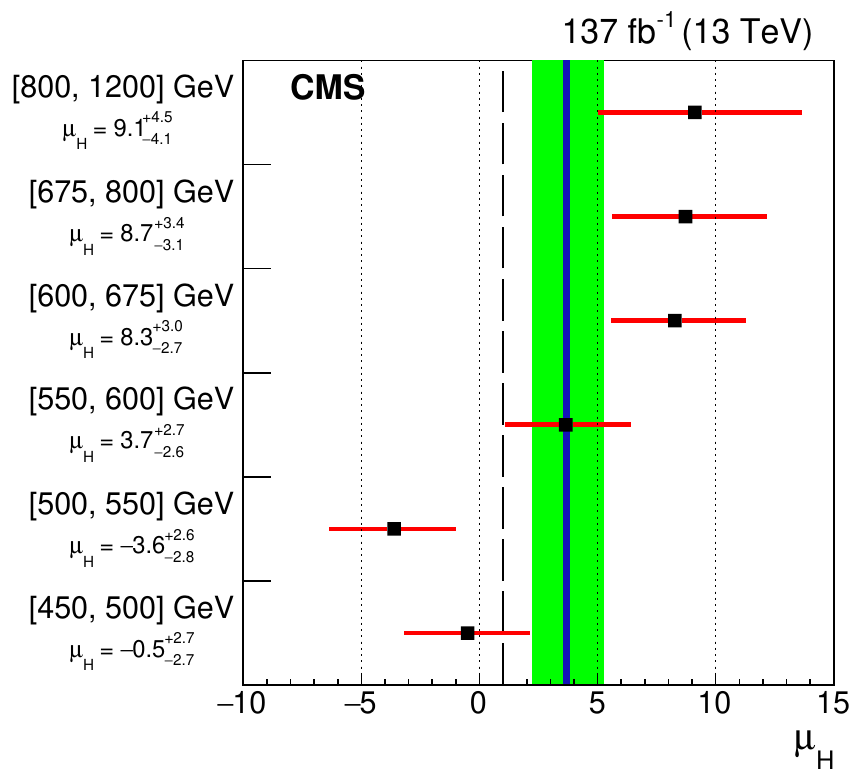}
\includegraphics[width=\cmsSmallFigWidth]{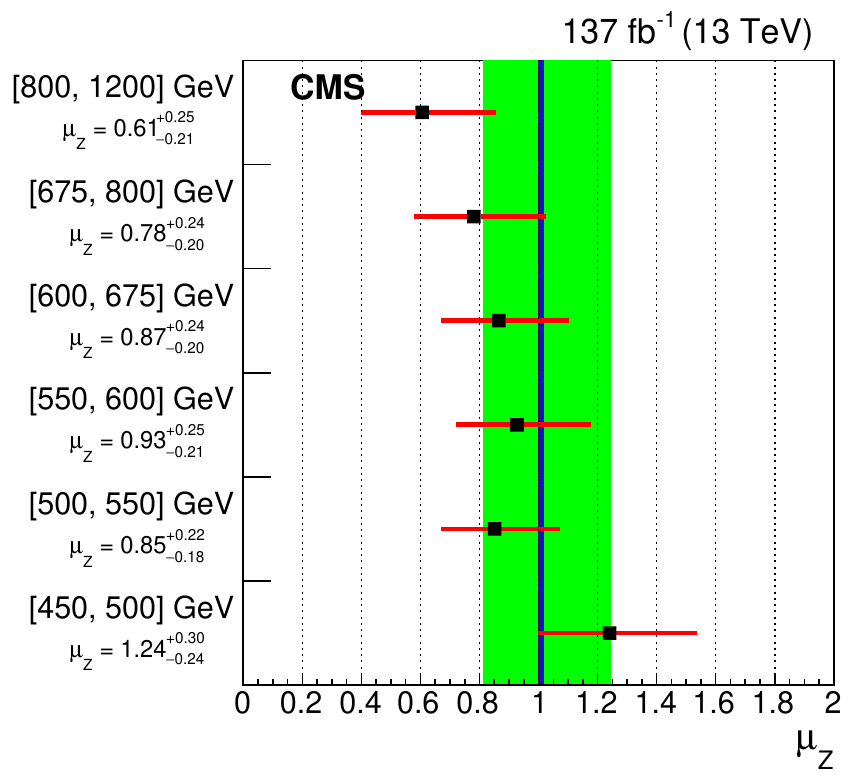}
\caption{
    The best-fit signal strength $\mu_{\PH}$ (black squares) and uncertainty (red lines) per \pt category based on the \HJMINLO~\cite{Hamilton:2012rf,Becker:2669113} prediction (left) and the same for $\mu_{\PZ}$ (right).
    The dashed black line indicates the SM expectation.
    The solid blue line and green band represents the combined best-fit signal strength and uncertainty, respectively, of $\mu_{\PH} =  \muValMINLO_{-\muErrLoMINLO}^{+\muErrHiMINLO}$ or $\mu_{\PZ}=\muZVal_{-\muZErrLo}^{+\muZErrHi}$ extracted from a simultaneous fit of all categories.
}
        \label{fig:ccc}
 \end{figure*}

To assess the compatibility between the observed signal strengths in the different jet \pt categories, an alternative fit to the data is performed.
In this fit, an independent $\mu_{\PH}$ is assigned to each of the six reconstructed jet \pt bins.
These signal strengths are unconstrained in the fit and are varied simultaneously.
All other parameters are profiled, as in the original fit.
Figure~\ref{fig:ccc} (left) illustrates the compatibility in the best fit signal strengths between the different \pt categories, showing an excess with respect to the SM expectation for categories with jet \pt above 550\GeV.
Separately, the same exercise is performed with an independent $\mu_{\PZ}$ in each \pt category.
The fitted signal strengths, shown in Fig.~\ref{fig:ccc} (right), are consistent with the SM expectation.

\begin{figure*}[hbtp]
  \centering
  \includegraphics[width=\cmsBigFigWidth]{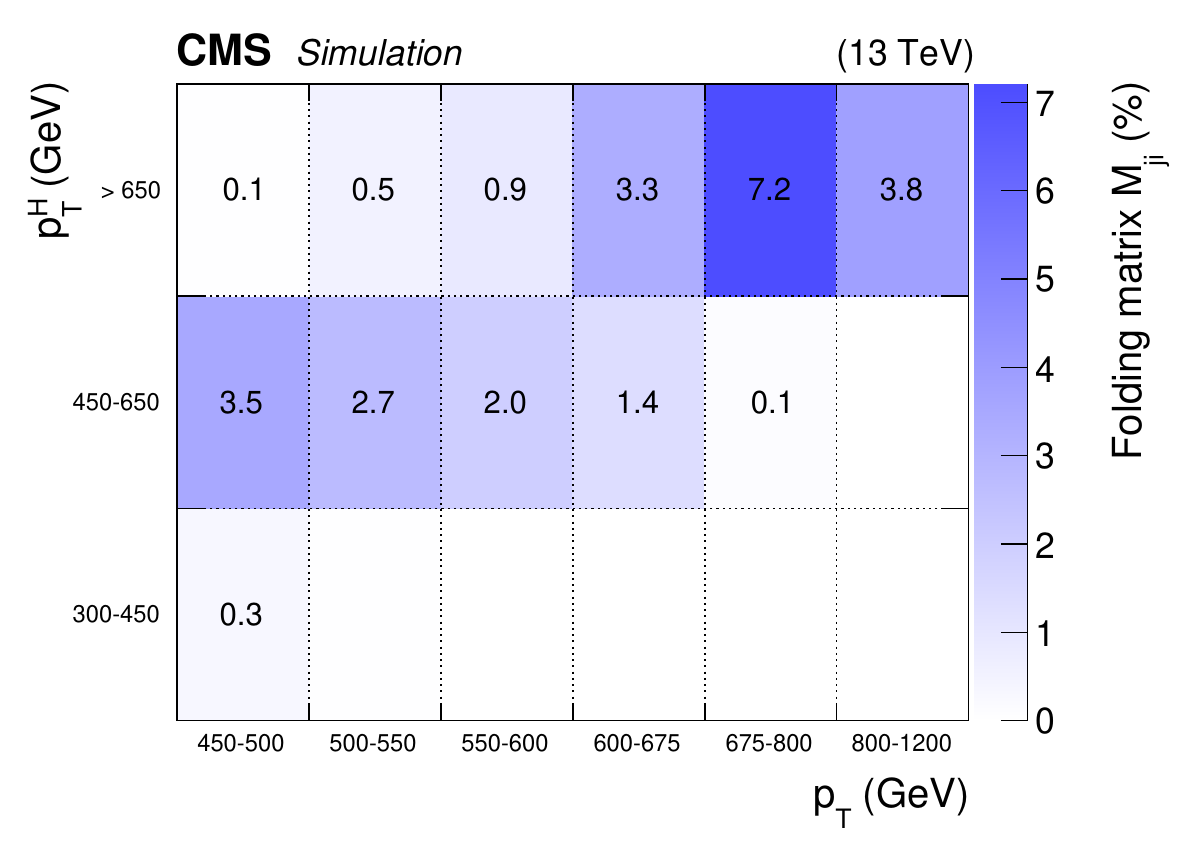}
  \caption{
    The folding matrix $M_{ji}$, defined as the product of the acceptance and the efficiency as a percentage for an $\PH(\bbbar)$ event in $\pt^{\PH}$ bin $j$ to be found in jet \pt bin $i$, for the {\ggH} \HJMINLO simulation.
  }
  \label{fig:response}
\end{figure*}

To facilitate comparisons with theoretical predictions,
we isolate and remove the effects of limited detector acceptance and response to the \ggH production cross section using a maximum-likelihood unfolding technique as described in Section~5 of Ref.~\cite{Sirunyan:2018sgc}.
In our treatment, the remaining Higgs boson production modes are assumed to occur at SM rates.
The \ggH signal is split into several bins according to the generated Higgs boson \pt ($\pt^{\PH}$), and each $\pt^{\PH}$ bin is considered as a separate process with a freely floating signal strength parameter in the likelihood model.
The respective $\pt^{\PH}$ bins are 300--450, 450--650, and $>$650\GeV.
This binning choice follows the simplified template cross section (STXS) recommendation~\cite{Berger:2019wnu}.
As the minimum reconstructed jet \pt is $450\GeV$, a negligible signal contribution is expected from events with $\pt^{\PH} < 300 \GeV$.
The folding matrix $M_{ji}$, defined as the product of the acceptance and the efficiency for an $\PH(\bbbar)$ event in $\pt^{\PH}$ bin $j$ to be found in jet \pt bin $i$, is shown in Fig.~\ref{fig:response} for the {\ggH} \HJMINLO simulation.
This matrix is found to be well-conditioned.
Therefore, we omit any regularization in the unfolding procedure~\cite{Hansen:LShape}.

The \ggH fiducial cross section in each STXS $\pt^{\PH}$ bin is then extracted by scaling the cross section found in simulation, imposing no selection requirements other than those on $\pt^{\PH}$, by the corresponding signal strength parameter.
The uncertainty in this value is taken from the correspondingly scaled signal strength uncertainty.
For the theoretical uncertainties, only those that affect the acceptance of signal events into the reconstructed selection are taken into account.
Based on the envelope of acceptance values from varying the renormalization and factorization scales by factors of two, this theoretical acceptance uncertainty is estimated to be 2\%.
We verify that this unfolding procedure is unbiased through signal injection studies.

\begin{figure}[hbtp]
  \centering
  \includegraphics[width=\cmsBigFigWidth]{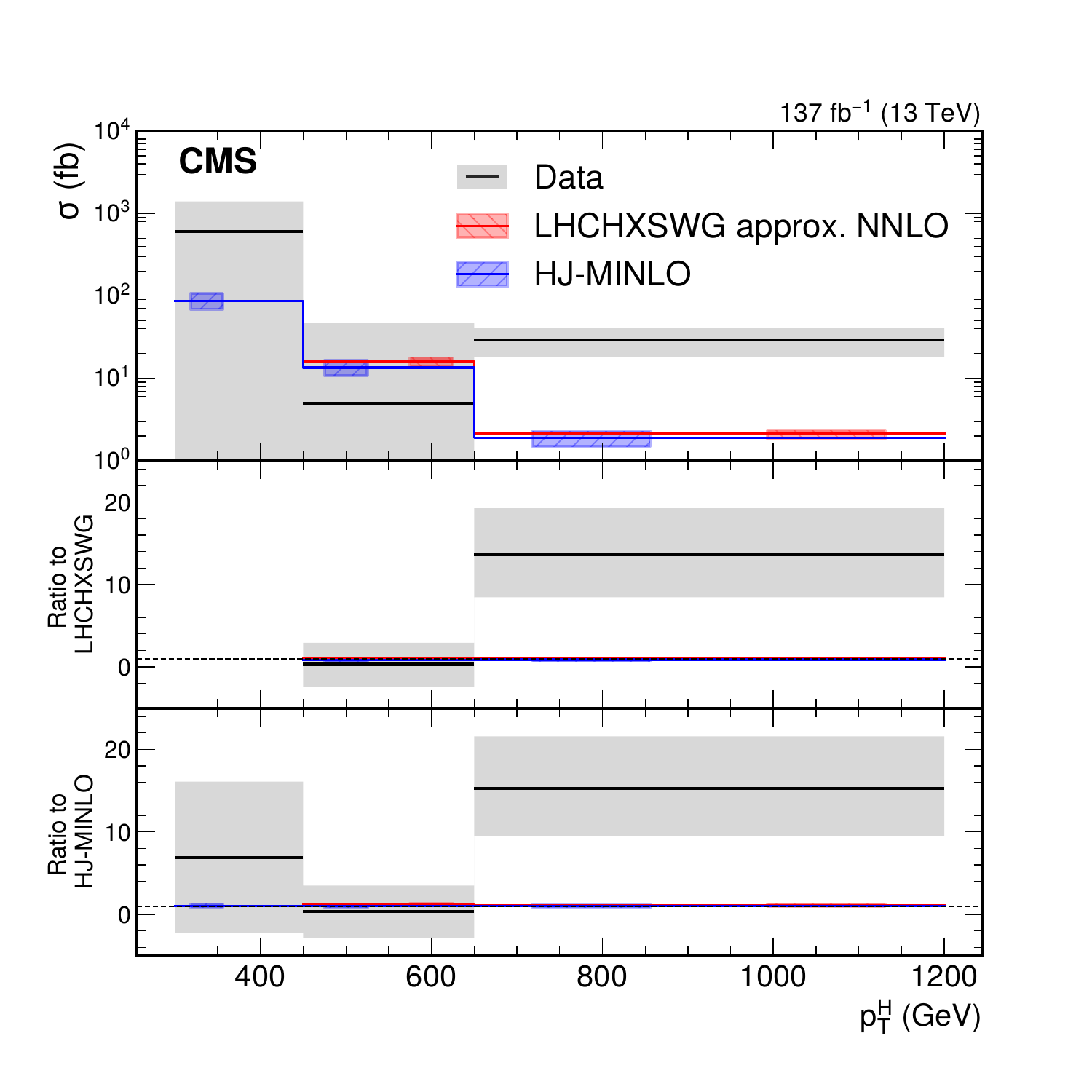}
  \caption{
    Measured \ggH differential fiducial cross section as a function of Higgs boson \pt shown in black, in comparison to the predictions of Ref.~\cite{Becker:2669113}, shown in red, and \HJMINLO~\cite{Hamilton:2012rf}, shown in blue.
    The two predictions are nearly identical.
    The larger gray band shows the total uncertainty in the measured cross section while the red and blue hatched bands show the uncertainties in the predictions of Ref.~\cite{Becker:2669113} and \HJMINLO, respectively.
    In the bottom two panels, the dotted line corresponds to a ratio of one.
    The relative uncertainties in the predictions of Ref.~\cite{Becker:2669113} and \HJMINLO are approximately 10 and 20\%, respectively.
  }
  \label{fig:fiducial}
\end{figure}

The result of this unfolding procedure is shown in Fig.~\ref{fig:fiducial} and Table~\ref{tab:XS}, along with the predicted cross sections from Ref.~\cite{Becker:2669113} and the predictions of the signal event generators described in Section~\ref{sec:samples}.
The correlation coefficients among the three $\pt^{\PH}$ bins are shown in Table~\ref{tab:covMatrix}.
The measured cross section uncertainty in the first $\pt^{\PH}$ bin is larger because of limited acceptance.
The first and second $\pt^{\PH}$ bins have a mild anti-correlation, primarily because of the imperfect jet energy response of the detector, which inflates the corresponding per-bin uncertainties in the unfolded cross section.
The observed cross section in the third $\pt^{\PH}$ bin has a smaller relative uncertainty than that in the second bin because of the larger magnitude of the central value in that bin.
With respect to the SM, the upward deviation of the cross section in the third $\pt^{\PH}$ bin, when profiling the other two, corresponds to a local significance of $2.6\,\sigma$.
When considering all three cross section parameters of interest simultaneously, the total deviation from the SM corresponds to a significance of $1.9\,\sigma$.

\begin{table*}[htb]
  \centering
  \topcaption{Measured and predicted \ggH differential fiducial cross section as a function of Higgs boson \pt. All cross sections are in units of \unit{fb}. The cumulative cross section predictions from Ref.~\cite{Becker:2669113} are converted to differential cross section predictions by subtraction assuming the cumulative cross section uncertainties are fully correlated.}
  \cmsTable{
  \begin{tabular}{lrlrlrl}
    $\pt^{\PH}$ (\GeVns{}) &  \multicolumn{ 2}{c}{300--450}   & \multicolumn{ 2}{c}{450--650}   & \multicolumn{ 2}{c}{$>$650}      \\\hline
    \multirow{ 2}{*}{Measured}  & $580$ & $\pm 790$ & $5$ & $\pm 43$   & $29$ & $\pm 11$  \\
              & & $\pm 720\stat \pm 350\syst $ & &  $\pm 37\stat \pm 22\syst$ & & $\pm 9\stat \pm 7\syst$  \\
    LHCHXSWG~\cite{Becker:2669113}  & \multicolumn{2}{c}{\NA} & $16.0$ & $^{+1.7}_{-2.0}$ & $2.1$ & $^{+0.2}_{-0.3}$  \\
    \HJMINLO~\cite{Hamilton:2012rf} & $89$ & $^{+20}_{-18} $   & $13.5$ &$^{+3.0}_{-2.7}$   & $1.9$ & $\pm 0.4$  \\
    Ref.~\cite{Sirunyan:2017dgc}    & $152$ & $\pm 46$         & $34$ & $\pm 10$        & $7.6$ & $\pm 3.0$  \\
  \end{tabular}}
  \label{tab:XS}
\end{table*}

\begin{table*}[htb]
  \centering
  \topcaption{Correlation coefficients between the three $\pt^{\PH}$ bins of the unfolded \ggH differential cross section measurement.}
  \begin{tabular}{lccc}
         $\pt^{\PH}$ (\GeV{}) &   300--450       & 450--650    & $>$650      \\    \hline
300--450  & $1.0$ &  $-0.18$ & $-0.002$ \\
450--650  & $-0.18  $ &  $1.0$ & $0.06$ \\
$>$650  & $-0.002$ &  $0.06$ & $1.0$ \\
  \end{tabular}
  \label{tab:covMatrix}
\end{table*}

\section{Summary}
\label{sec:summary}

An inclusive search for the standard model (SM) Higgs boson decaying to a bottom quark-antiquark pair and reconstructed as a single large-radius jet with transverse momentum $\pt > 450 \GeV$ has been presented.
The search uses a data sample of proton-proton collisions at $\sqrt{s}=13\TeV$, corresponding to an integrated luminosity of \analysisLumi.
The associated production of a \PZ boson and jets is used to validate the method and is measured to be consistent with the SM prediction.
The inclusive Higgs boson signal strength is measured to be $\mu_{\PH} = \muValMINLO\pm\muStatErrMINLO\stat_{-\muSystErrLoMINLO}^{+\muSystErrHiMINLO}\syst_{-\muThyErrLoMINLO}^{+\muThyErrHiMINLO}\thy = \muValMINLO_{-\muErrLoMINLO}^{+\muErrHiMINLO}$, based on the theoretical prediction from the \HJMINLO generator for the gluon fusion production mode.
The measured $\mu_{\PH}$ corresponds to an observed significance of $\muObsSigMINLO$ standard deviations ($\sigma$) with respect to the background-only hypothesis, while the expected significance of the SM signal is $\muExpSigMINLO\,\sigma$.
The significance of the observed excess with respect to the SM expectation is $\muObsSigMINLOMuOne\,\sigma$.
With respect to the previous CMS result, the relative precision of the $\mu_{\PH}$ measurement improves by approximately a factor of two because of the increased integrated luminosity, an improved \PQb tagging technique based on a deep neural network, and smaller theoretical uncertainties.
Finally, the differential cross section for the \pt of a Higgs boson produced through gluon fusion, assuming the other production modes occur at the SM rates, in the phase space regions recommended by the LHC simplified template cross section framework has also been presented.
An excess is seen for Higgs boson $\pt>650\GeV$ with a local significance of $2.6\,\sigma$ with respect to the SM expectation including the Higgs boson.

\begin{acknowledgments}
  We congratulate our colleagues in the CERN accelerator departments for the excellent performance of the LHC and thank the technical and administrative staffs at CERN and at other CMS institutes for their contributions to the success of the CMS effort. In addition, we gratefully acknowledge the computing centers and personnel of the Worldwide LHC Computing Grid for delivering so effectively the computing infrastructure essential to our analyses. Finally, we acknowledge the enduring support for the construction and operation of the LHC and the CMS detector provided by the following funding agencies: BMBWF and FWF (Austria); FNRS and FWO (Belgium); CNPq, CAPES, FAPERJ, FAPERGS, and FAPESP (Brazil); MES (Bulgaria); CERN; CAS, MoST, and NSFC (China); COLCIENCIAS (Colombia); MSES and CSF (Croatia); RIF (Cyprus); SENESCYT (Ecuador); MoER, ERC IUT, PUT and ERDF (Estonia); Academy of Finland, MEC, and HIP (Finland); CEA and CNRS/IN2P3 (France); BMBF, DFG, and HGF (Germany); GSRT (Greece); NKFIA (Hungary); DAE and DST (India); IPM (Iran); SFI (Ireland); INFN (Italy); MSIP and NRF (Republic of Korea); MES (Latvia); LAS (Lithuania); MOE and UM (Malaysia); BUAP, CINVESTAV, CONACYT, LNS, SEP, and UASLP-FAI (Mexico); MOS (Montenegro); MBIE (New Zealand); PAEC (Pakistan); MSHE and NSC (Poland); FCT (Portugal); JINR (Dubna); MON, RosAtom, RAS, RFBR, and NRC KI (Russia); MESTD (Serbia); SEIDI, CPAN, PCTI, and FEDER (Spain); MOSTR (Sri Lanka); Swiss Funding Agencies (Switzerland); MST (Taipei); ThEPCenter, IPST, STAR, and NSTDA (Thailand); TUBITAK and TAEK (Turkey); NASU (Ukraine); STFC (United Kingdom); DOE and NSF (USA).

  \hyphenation{Rachada-pisek} Individuals have received support from the Marie-Curie program and the European Research Council and Horizon 2020 Grant, contract Nos.\ 675440, 752730, and 765710 (European Union); the Leventis Foundation; the A.P.\ Sloan Foundation; the Alexander von Humboldt Foundation; the Belgian Federal Science Policy Office; the Fonds pour la Formation \`a la Recherche dans l'Industrie et dans l'Agriculture (FRIA-Belgium); the Agentschap voor Innovatie door Wetenschap en Technologie (IWT-Belgium); the F.R.S.-FNRS and FWO (Belgium) under the ``Excellence of Science -- EOS" -- be.h project n.\ 30820817; the Beijing Municipal Science \& Technology Commission, No. Z191100007219010; the Ministry of Education, Youth and Sports (MEYS) of the Czech Republic; the Deutsche Forschungsgemeinschaft (DFG) under Germany's Excellence Strategy -- EXC 2121 ``Quantum Universe" -- 390833306; the Lend\"ulet (``Momentum") Program and the J\'anos Bolyai Research Scholarship of the Hungarian Academy of Sciences, the New National Excellence Program \'UNKP, the NKFIA research grants 123842, 123959, 124845, 124850, 125105, 128713, 128786, and 129058 (Hungary); the Council of Science and Industrial Research, India; the HOMING PLUS program of the Foundation for Polish Science, cofinanced from European Union, Regional Development Fund, the Mobility Plus program of the Ministry of Science and Higher Education, the National Science Center (Poland), contracts Harmonia 2014/14/M/ST2/00428, Opus 2014/13/B/ST2/02543, 2014/15/B/ST2/03998, and 2015/19/B/ST2/02861, Sonata-bis 2012/07/E/ST2/01406; the National Priorities Research Program by Qatar National Research Fund; the Ministry of Science and Higher Education, project no. 02.a03.21.0005 (Russia); the Tomsk Polytechnic University Competitiveness Enhancement Program and ``Nauka" Project FSWW-2020-0008 (Russia); the Programa Estatal de Fomento de la Investigaci{\'o}n Cient{\'i}fica y T{\'e}cnica de Excelencia Mar\'{\i}a de Maeztu, grant MDM-2015-0509 and the Programa Severo Ochoa del Principado de Asturias; the Thalis and Aristeia programs cofinanced by EU-ESF and the Greek NSRF; the Rachadapisek Sompot Fund for Postdoctoral Fellowship, Chulalongkorn University and the Chulalongkorn Academic into Its 2nd Century Project Advancement Project (Thailand); the Kavli Foundation; the Nvidia Corporation; the SuperMicro Corporation; the Welch Foundation, contract C-1845; and the Weston Havens Foundation (USA).

\end{acknowledgments}

\bibliography{auto_generated}
\cleardoublepage \appendix\section{The CMS Collaboration \label{app:collab}}\begin{sloppypar}\hyphenpenalty=5000\widowpenalty=500\clubpenalty=5000\vskip\cmsinstskip
\textbf{Yerevan Physics Institute, Yerevan, Armenia}\\*[0pt]
A.M.~Sirunyan$^{\textrm{\dag}}$, A.~Tumasyan
\vskip\cmsinstskip
\textbf{Institut f\"{u}r Hochenergiephysik, Wien, Austria}\\*[0pt]
W.~Adam, F.~Ambrogi, T.~Bergauer, M.~Dragicevic, J.~Er\"{o}, A.~Escalante~Del~Valle, R.~Fr\"{u}hwirth\cmsAuthorMark{1}, M.~Jeitler\cmsAuthorMark{1}, N.~Krammer, L.~Lechner, D.~Liko, T.~Madlener, I.~Mikulec, F.M.~Pitters, N.~Rad, J.~Schieck\cmsAuthorMark{1}, R.~Sch\"{o}fbeck, M.~Spanring, S.~Templ, W.~Waltenberger, C.-E.~Wulz\cmsAuthorMark{1}, M.~Zarucki
% ...
\vskip\cmsinstskip
\dag: Deceased\\
1:  Also at Vienna University of Technology, Vienna, Austria\\
% ...
\end{sloppypar}
\end{document}